\documentclass{amsart}
\usepackage{epsfig}
\usepackage{oldlfont}
\usepackage{citesort}

\newtheorem*{ack}{Acknowledgment}

\newcommand{\ket}[1]{\mid #1 \rangle}
\newcommand{\braket}[2]{\langle #1 \mid #2 \rangle}

\newcommand{\Ntil}{ N \! \! \! \! \! {}^{}_{{}_{\sim}}}

\newcommand{\pic}[5]{\raisebox{#3pt}
{\hspace{#4pt} \epsfig{file=figs/#1.ps,height=#2pt,silent=} 
\hspace{#5pt}}}
\newcommand{\kd}[1]{\mathchoice{
\pic{#1}{40}{-18}{-1}{2}}{
\pic{#1}{12}{-3}{1}{1}}{
\pic{#1}{9}{-2}{-3}{1}}{
\pic{#1}{7}{-1}{-1}{0}}}

\newcommand{\skd}[1]{\mathchoice{
\pic{#1}{24}{-8}{-1}{2}}{
\pic{#1}{20}{-5}{1}{1}}{
\pic{#1}{20}{-10}{-3}{1}}{
\pic{#1}{18}{-8}{-1}{0}}}

\begin{document}

\title{Quasilocal Energy for Spin-net Gravity}

\author{Seth A. Major}

\date{December 1999}
\address{Institut f\"ur Theoretische Physik 
\\ Der Universit\"at Wien \\ Boltzmanngasse 5\\
A-1090 Wien AUSTRIA}
\email{Updated address: smajor1@swarthmore.edu}

\begin{abstract}
	The Hamiltonian of the gravitational field defined in a bounded
	region is quantized.  The classical Hamiltonian, and starting
	point for the regularization, is a boundary term required by
	functional differentiability of the Hamiltonian constraint.  It is
	the quasilocal energy of the system and becomes the ADM mass in
	asymptopia.  The quantization is carried out within the framework
	of canonical quantization using spin networks.  The result is a
	gauge invariant, well-defined operator on the Hilbert space
	induced by the state space on the whole spatial manifold.  The
	spectrum is computed.  An alternate form of the operator, with the
	correct naive classical limit, but requiring a restriction on the
	Hilbert space, is also defined.  Comparison with earlier work and
	several consequences are briefly explored.
\end{abstract}

\maketitle

\begin{flushright}
		UWThPh-1999-40
\end{flushright}

\section{Introduction}

Fundamental physical observables of the gravitational field, such as
energy and angular momentum, are notoriously hard to define
satisfactorily.  In asymptopia, where there exist the well-defined
Bondi \cite{bondi,sachs} and Arnowitt-Deser-Misner (ADM) \cite{ADM}
masses, the quantities are non-local.  The inherent non-locality is
usually attributed to the equivalence principle; one needs at least
two observers to distinguish geodesic deviation from
acceleration-dependent quantities.  Nevertheless, given the existence
of the Bondi and ADM energy definitions on 2-surfaces at null and
spatial infinities, it is reasonable to expect a degree of
localization is possible.  One possibility given by Penrose
\cite{penroseQE} is to associate quasilocal quantities to compact,
oriented spatial 2-surfaces.  Given the difficulty in defining
gravitational observables, it is perhaps not surprising that there
exists a plethora of proposed quasilocal quantities (\cite{hawking} -
\cite{lau}).  The definitions of quasilocal energy may be
distinguished by the ability to satisfy various criteria.  These
include vanishing in flat space, taking reasonable values for
spherically symmetric solutions, and approaching the ADM and Bondi
values in appropriate limits.  It is by no means simple to satisfy
these criteria \cite{B}, \cite{hayward}.

With the complications involved in defining energy even in classical
general relativity, it may seem perilous to define quantum quasilocal
energy.  However, in the context of the canonical ($3+1$) theory there
is a clear selection criteria for the correct physical quantity.  When
defined within a bounded region, functional differentiability of the
constraints requires that, generically, surface terms must be added to
make the theory well-defined.  The surface term required by the
Hamiltonian constraint is the Hamiltonian for the system on shell. 
In general, the method of functional differentiability generates
boundary conditions on the phase space variables, gauge parameters,
and surface terms (as is explored in some depth in Ref.  \cite{HM} for
gravity and BF theory).

Originally, it was noticed that a surface term had to be added to the 
action so as to match the asymptotic expression.  However, Regge and 
Teitelboim found that gravitational theory was simply inconsistent 
without this boundary term \cite{RT}. The gravitational 
action has derivatives of the phase space functions, so the 
variation generates surface terms.  It is only when these terms vanish 
(or are canceled) that the theory has a well-defined variational 
principle.  The boundary Hamiltonian or quasilocal energy is of this 
form; it cancels a term arising in the variation of the Hamiltonian 
constraint.

This paper provides a quantization of the surface observable
\begin{equation}
	\label{hclass}
H_{\partial \Sigma}(N) = 
\frac{1}{4\pi G} \int_{\partial \Sigma} d^{2}x \epsilon^{ijk} N n_{a}
A_{b}^{i}E^{bj}E^{ak}.
\end{equation}
This reduces to the quasilocal energy of Brown and York \cite{BY}, to 
the ADM energy in asymptopia, and to the Misner-Sharp mass in 
spherical symmetry (\cite{HM}).

The quantization is carried out in the context of canonical quantum 
gravity in the real new variables \cite{NV,barbero} or, succinctly, 
``spin-net gravity''\footnote{A word on the choice of name: ``Spin-net 
gravity'' is chosen to reflect the current state of affairs.  It is 
widely recognized that the kinematic state space of the canonical 
quantization of gravity in terms of real connection variables has a 
basis in terms of spin networks.  The descriptive name, spin-net 
gravity, emphasizes the fundamental importance of spin networks to the 
theory both as a kinematic state space (in which case one might say 
that one has ``quantum geometry'') and the very critical assumption 
that spin networks also serve to describe the dynamics. There are a 
number of results which suggest that spin networks fill this role -- 
as this paper does -- but the issue is not resolved.} in which the 
state space is built from functions of holonomies based on graphs.  
(See Ref.  \cite{CRrev} for a recent review.)  Perhaps the most 
remarkable result of this study is the discreteness in geometric 
operators.  Length \cite{Tlength}, area (\cite{RSareavol,QGII,FLR}), 
volume (\cite{RSareavol,lollvol,Tvol,BorissovetalMatrix,QSDII,Lvol}), 
and angle \cite{SAMangle} have been found to have fully discrete 
spectra.  In this approach to quantum gravity space is discrete.

There have been two previous quantizations of Eq.  (\ref{hclass}). 
The first, by Baez, Muniain, and Piriz, was completed before many of
the spin-net techniques were developed \cite{BMP}.  The action of the
operator was qualitatively described as a ``shift.''  In the second,
two ADM energy operators were introduced in the framework of
Thiemann's quantum spin dynamics (\cite{QSDI} - \cite{QSDVI}).  The
classical expression used in that quantization was different than the
one used here.  In fact, the expression is only weakly equal to Eq. 
(\ref{hclass}) \cite{Tasym}.  Nevertheless, the resulting quantum
operator is very similar and reduces to this ADM expression under the
same restrictions.  A more extensive comparison between the operators
is given in Section \ref{conclusion}.  The present work completes the
quantization of the boundary Hamiltonian, even in a non-asymptotically
flat context.

The quantum definition is to satisfy modest criteria.  I only ask that
the operator be well-defined on the Hilbert space induced from the
full gauge invariant space.  Important criteria, including the proper
behavior on semiclassical states and the appropriate algebra of
boundary observables, is left for further investigation.  Definitions
of simple semiclassical states or weaves \cite{weave} are explored in
\cite{BM}.

The remainder of this paper is organized as follows: Spin-net gravity 
is briefly reviewed in Section \ref{qg}.  This serves to base the 
definition of the quasilocal energy in the framework of the classical 
theory and provides an opportunity to fix notation and units.  The 
main developments are contained in Section \ref{qe} where the operator 
is regularized in \ref{reg} and given final form in 
\ref{recouplingnote}.  An alternate quantization is given in 
\ref{qevol} and the full spectra of the resulting operators is 
presented in \ref{spectra}.  This paper concludes with a comparison 
with Thiemann's $E_{ADM}$ operator and some discussion on wider 
implications of the quantum quasilocal energy operator.

\section{The setting: Canonical Quantum Gravity}
\label{qg}

This section sets the classical framework for the operator definitions 
and provides the basic elements of spin-net quantization.  Meanwhile, I 
use the opportunity to fix signs, factors, and units.  For an 
introduction, rather than a brief review, the reader is encouraged to 
read Refs. \cite{CRrev,QGI,QGII,QGIII,SAMsn,roberto,RC}.

For four-dimensional spacetimes $M$ with $M=\Sigma \times {\mathbb R}$, 
where $\Sigma$ is compact, the ($3+1$)-action for vacuum, Riemannian 
general relativity is
\begin{equation}
	\label{action}	
	S[E^{ai}, A_a^i; \Lambda^i, N^a, N]  
	= { 1 \over 8 \pi G } \int_{t_1}^{t_2} dt \int_{\Sigma} d^3x 
	[\ E^{ai}\dot{A}_a^i - N{\cal H} - N^a{\cal D}_a -\Lambda^i G^i\ ].
\end{equation}
The action is written in terms of a real, $su(2)$-valued connection 
one-form $A_a^i$ and its momenta, a densitized inverse triad $E^{ai}$.  
I have included the overall factor depending on Newton's constant $G$ 
($c=1$ throughout the paper).  The phase space variables 
satisfy\footnote{As Immirzi has emphasized, in the canonical 
transformation used to define the connection there is a family of 
choices generated by one non-zero, real parameter $\gamma$, 
${}^{\gamma}A_{a}^{i} = \Gamma_{a}^{i} - \gamma K_{a}^{i}, 
{}^{\gamma}E^{ai} = (1/ \gamma)E^{ai}$ \cite{I}.  I take $\gamma =1$ 
until the final section, when it is included in the spectra of the 
quasilocal energy.}
\begin{equation}
\label{PB}
\left\{ A_a^i(x), E^{b}_j(y) \right\} = 8 \pi G \, \delta_a^b 
\delta^i_j 
\delta(x,y). 
\end{equation}
The Lagrange multipliers $N$, $N^a$, and $\Lambda^i$ are, 
respectively, the lapse, shift, and $SU(2)$ gauge rotation parameters.  
Varying the action with respect to the last two functions gives the 
constraints
\begin{equation}
	\begin{split}
		\label{gauss}
		G^i &\equiv D_aE^{ai} \approx 0 \\
	{\cal D}_a &\equiv E^{bi}\partial_a A_b^i - \partial_b(E^{bi}A_a^i) 
	\approx 0 
	\end{split}
\end{equation}
where  $D_a \lambda^i = \partial_a \lambda^i + \epsilon^{ijk}
A_a^j \lambda^k$.  Varying the lapse gives the Hamiltonian constraint. 
Defining
$F_{ab}^i = \partial_{[a} A_{b]}^i + \epsilon^{ijk}A_a^j A_b^k$,
one may express the constraint in integrated form as \cite{barbero}
\begin{equation}
	\label{ham}
	H(N) = \frac{1}{8 \pi G}
	\int_\Sigma d^3x \, N \left( \epsilon^{ijk} E^{ai} E^{bj} F_{ab}^k
	- 4 E^{a}_{[i} E_{j]}^{b} ( A_{a}^{i} - \Gamma_{a}^{i} )
	( A_{b}^{j} - \Gamma_{b}^{j} ) \right) \approx 0.
\end{equation}
Due to the choice of using real variables, the additional term has 
been added to the constraint \cite{barbero}.  The triad-compatible 
connection $\Gamma_{a}^{i}$ satisfies $ {\cal D}_{a} E^{bi} = 
\partial_{a} E^{bi} +\epsilon^{ijk} \Gamma_{a}^{j} E^{bk} + 
\Gamma_{ac}^{b} E^{ci} - \Gamma_{ac}^{c} E^{bi} = 0$.  This constraint 
generates time evolution.  A key observation which affects the 
definition of the quasilocal energy is that the Hamiltonian constraint 
has density weight +2 so the lapse function has density weight -1.

As bounded spatial regions are the subject of this work, it is best to 
start by fixing notation.  The calculations are in a spacetime of the 
form $M = \Sigma \times {\mathbb R}$, although the space has at least 
one compact subset $I \subset \Sigma$ such that the boundary of the 
interior $I$, $\partial I$, is homeomorphic to a 2-sphere.  I usually 
denote the boundary $\partial I$ as the surface $S$.  The topology of 
$\Sigma$ is not specified but two examples are worth keeping in mind.  
One is a compact spatial slice $\Sigma$ with one boundary and two 
``interiors'' $I$ and its complement $I^{*}$.  The other is 
the topology of the asymptotically flat spacetimes in which $\Sigma$ 
is homeomorphic to ${\mathbb R}^{3}$ with a compact ball cut out.

When the theory of Eq.  (\ref{action}) is applied to a bounded 
spatial region $I$, it is no longer well-defined \cite{HM}.  The 
problem arises because the phase space of the compact theory 
does not contain the physical solutions of the bounded one \cite{RT}; 
the variational problem has no solutions.  This comes about as 
boundary terms arise in the variation, making Hamilton's equations ill 
defined; dynamics takes the system outside of the phase space.

The ($3+1$)-action, or the constraints, must be functionally 
differentiable.  When the theory is defined in a finite region, this 
requires the addition of surface terms and/or the imposition of 
boundary conditions.  These surface terms, without which the theory 
would be inconsistent, are the fundamental observables and necessarily 
satisfy the same algebra as the constraints \cite{HM}.

Under the variation of the connection, the Hamiltonian constraint 
generates a surface term which must either vanish through the 
imposition of boundary conditions or be canceled by another surface 
term.  If the lapse is non-vanishing on $\partial I$ then the surface 
term of Eq.  (\ref{hclass}) must be added to the constraint.  It is 
this term which is the quasilocal energy and reduces to the ADM energy 
in asymptopia.  To ensure that the theory is dynamically well-defined 
one must impose a complete set of conditions on the boundary $\partial 
I$.  To be concrete, I consider the specific boundary conditions:
\begin{equation}
	\label{bcs}
	\begin{split}
		&\delta E^{ai}|_{\partial I} = 0 \text{ which includes fixing 
		the ``area density'' } n_{a} \delta E^{ai}|_{\partial I} = 0 \\
		&\delta N^{a}|_{\partial I} =0; \; \delta N|_{\partial I} =0
	\end{split}
\end{equation}
See Ref. \cite{HM} (or, for a more general setting, Ref.  \cite{dis}) 
for details.

This work is devoted to quantizing the observable of Eq.  
(\ref{hclass}).  Before beginning this, I review the 
quantization program in which the quasilocal energy operation is 
defined.  The mathematically precise formulation is developed in Refs.  
\cite{AI,ALproject1,ALrep,ALProj,MM,BS,ALMMT}.

The natural framework for a diffeomorphism invariant gauge theory is 
based on Wilson loops \cite{JS,RSloop}.  However, to build 
three-dimensional geometry it is necessary to use a more general 
structure based on graphs.  The quantum configuration space is most 
appropriately constructed from holonomies along the edges of a graph.  
I denote a graph embedded in $\Sigma$ by ${\mathsf G}$.  It contains a 
set of $N$ edges ${\bf e}$ and a set of vertices ${\bf v}$.  The key 
idea is that every smooth connection $A$ associates a group element to 
an edge $e$ of $\bf e$ via the holonomy,
\[
U_e (A) := {\cal P} \exp  \int_e dt \dot{e}^a A_a (e(t)).
\]
Here, $A_{a} := A_{a}^{i} \tau^{i}$ with $\tau^{i}$ proportional to
the Pauli matrices via $\tau^{i} = - \tfrac{i}{2} \sigma^{i}$.  Given
an embedded graph, holonomies along the edges, and a complex
(Haar-integrable) function on $SU(2)^{N}$, one may define a
``cylindrical function''
\[
\Psi_{{\mathsf G}, f}(A) = f\left( U_{e_{1}}(A), \dots, U_{e_{N}}(A) 
\right)
\]
(The function $\Psi$ is ``cylindrical'' since $f$ only depends on a
finite number of directions in the space of connections; it is
constant on all other directions.)  With the appropriate norm, the
completion of this function space gives the Hilbert space ${\cal H}$.

A basis on ${\cal H}$ is given by spin network states \cite{baez}.  I
denote a spin network ${\cal N}$ by the triple $(\mathsf{G}; {\bf i,
n})$ of an oriented graph $\mathsf{G}$, labels on the vertices (or
``intertwiners''), ${\bf i}$, and integer edge labels, ${\bf n}$,
indexing the representation carried by the edge.  The corresponding
spin net state $\ket{s}$ in $\cal H$ is defined in the connection
representation as
\[
\braket{A}{s} 
\equiv \braket{A}{ {\mathsf G} \, {\bf  i \,  n} } 
:= \prod_{v \in {\bf v}( {\mathsf G} ) }
{\bf i}_v  \circ \otimes_{e \in e({\mathsf G})} U_{e}^{(n_{e})}[A]
\]
where the holonomy $U_{e}^{(n_{e})}$ along edge $e$ is in the $n_{e}/2$ 
irreducible representation of $SU(2)$.  When the intertwiners ``tie 
up'' all the incident edges -- when they are invariant tensors on the 
group -- these states are gauge invariant.

As the triads are dual to pseudo two-forms, they most comfortably live 
on 2-surfaces, denoted by $S$.
\begin{equation}
\label{Edef}
E_{S}^i = \int_S d^2 \sigma \, n_a(\sigma) E^{ai} (x(\sigma))
\end{equation}
in which $\sigma$ are coordinates on the surface and $n_a = 
\epsilon_{abc} \tfrac{dx^b}{d\sigma_1}\tfrac{dx^c}{d\sigma_2}$ is the 
normal.  The action of the triads on a function of holonomies may be 
computed from the Poisson brackets, Eq.  (\ref{PB}).  A short 
calculation shows that the bracket with a single edge $U_{e}$ is given 
by
\begin{equation}
	\label{triadbracket}
	\left\{ E^{ai}(x) , U_{e}(A) \right\} = 
	- 8 \pi G \int_{e} dt \, \dot{e}^{a}(t) \, \delta(x, e(t)) 
	\, J^{i}_{(e)} \cdot U_{e}(A)
\end{equation}
in which $J^{i}_{(e)} \cdot U_{e}$ denotes the action of the left (or 
right) invariant vector fields on the group element 
$U_{e}$.\footnote{While $J_{(e)}^{i}$ does satisfy the algebra of Eq.  
(\ref{Jalg}), it has an additional property: under orientation 
reversal of the edge $e$, $J_{(e)}^{i}$ changes sign \cite{QGII}.  
This may be seen directly from Eq.  (\ref{triadbracket}).} The bracket 
will generally have a sum of terms.  When a cylindrical function based 
on a graph ${\mathsf G}$, say $c_{{\mathsf G}}(A)$, is used, the triad 
defined on a surface, $E_{S}^{i}$, gives a sum over all intersections, 
$v$, of the surface and the graph ${\mathsf G}$
\[
\left\{ E_{S}^{i}, c_{{\mathsf G}}(A) \right\} 
= - \frac{8 \pi G}{2} \sum_{v \in  S \cap {\mathsf G} } 
\sum_{I \dashv v} \chi^{S}_{I} J_{I}^{i} \cdot c_{{\mathsf G}}.
\]
The sum is over all edges $I$ incident to $v$ (denoted $I \dashv v$).  
The geometric factor $\chi^{S}_{I}$ is defined by
\begin{equation}
	\label{chi}
	\chi^{S}_{I} =
	\begin{cases}
		+1 & \text{when the orientation of $e_{I}$ is aligned with $n_{a}$}\\
		0 &  \text{when the edge is tangent}\\
		-1 & \text{when the orientation of $e_{I}$ is anti-aligned with 
		$n_{a}$}.
	\end{cases}
\end{equation}
This geometrical factor ties the tangent space of the edge (through 
$\dot{e}_{I}$) to the orientation of the surface.  Due to the behavior 
of $\hat{J}_{(e)}^{i}$ under edge orientation reversal, this factor 
maintains edge orientation independence and surface orientation 
dependence.

There are two further remarks to make.  First, the result is 
non-vanishing only when there is at least one intersection between the
graph ${\mathsf G}$ and the surface $S$.  Second, the overall factor of
$\tfrac{1}{2}$ can be seen to arise from a ``thickened surface'' 
regularization \cite{QGI}. 

With this preparation, one may define the quantum triad 
operator on a spin network state $\ket{s} = \ket{{\mathsf G} \, {\bf i 
\, n}}$
\begin{equation}
	\label{Eop}
	\hat{E}_{S}^{i} \ket{s} 
	:=  - (4 \pi G)^{2} 
	\sum_{v \in S \cap {\mathsf G}} 
	\sum_{I \dashv v}
	\chi^{S}_{I} \hat{J}_{I}^{i} \ket{s}
\end{equation}
where the angular momentum-like operator $\hat{J}_{(e)}^{i} \equiv i \hbar 
J_{(e)}^{i}$ satisfies the usual algebra
\begin{equation}
	\label{Jalg}
	\left[ \hat{J}_{(e)}^{i}, \hat{J}_{(e')}^{j} \right] 
	= i\, \hbar 
	\, \epsilon^{ijk} \delta_{e , e'} \hat{J}_{(e)}^{k}. 
\end{equation}
(The $\delta$-function restricts the relation to one edge; $\hat{J}^{i}$ 
on distinct edges commute.)  The triad operator of Eq.  (\ref{Eop}) is 
essentially self-adjoint in the Hilbert space of quantum gravity 
\cite{QGI,ALProj}.  The diagrammatic form of this operator is the 
``one-handed'' \cite{SAMsn}
\begin{equation}
	\label{Ediagram}
	\hat{E}_{S}^{i} \ket{s} 
	= - i l^{2} 
	\sum_{v \in S \cap {\mathsf G}} 
	\sum_{I \dashv v}
	\chi^{S}_{I} 
	\pic{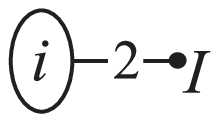}{14}{-3}{-1}{2} \ket{s}
\end{equation}
in which the index $i$ is the internal space index.  The length scale 
of the theory, $l$, is defined by $l^{2} = 4 \pi \hbar G$.  The 
grasping is chosen such that, in the plane of the diagram, the 2-line 
is on the left when the orientation on the edge points up (vice versa 
for downward orientations) \cite{RC,roberto}.  Though the overall action is 
orientation independent, such a diagrammatic representation of 
grasping involves the choice of a sign.  The geometric factor 
$\chi^{S}_{I}$ in the expression for the ``unclasped hand'' in 
Eq.  (\ref{Ediagram}) ensures that the operator is independent of 
edge orientation \cite{roberto}.

The quantum configuration space based on graphs and triads which act as 
signed angular momentum operators are the basic elements of spin-net 
gravity.  The definition of the quasilocal energy makes use of two 
further results, the geometric operators of area and volume.

The area of a surface $S$ may be, for simplicity, specified by $z=0$ 
in an adapted coordinate system.  Expressed in terms of the triad 
$E^{ai}$, the area of the surface only depends on the $3$rd vector 
component via \cite{RSareavol,FLR,QGI}
\begin{equation}
	\label{areae}
	A_{S} = \int_{S} d^{2}x \sqrt{E^{3i}  E^{3i}}.
\end{equation}
The quantum operator is defined using the operators of Eq.  
(\ref{Eop}) and by partitioning the surface $S$ so that only one edge 
or vertex threads through each cell in the partition.  The integral of 
Eq.  (\ref{areae}) becomes a sum over operators that act only at 
intersections of the surface with the spin network
\begin{equation}
	\label{areaop}
	\hat{A}_{S} \ket{s} = (4 \pi G)^{2} \sum_{v} \sum_{I,J} 
	\sqrt{ \chi^{S}_{I} \chi^{S}_{J} \hat{J}_{I} \cdot \hat{J}_{J}}.
\end{equation}
The spectrum may be computed with recoupling theory \cite{FLR} or 
operator methods \cite{QGI}.  In both cases it is found by considering 
all the intersections of the spin network with the surface $S$, 
including vertices which lie in the surface.  The edges incident to a 
vertex in the surface may be divided into three categories: those 
which have tangents aligned with the surface normal $j_{p}$, 
anti-aligned $j_{n}$, and tangent to the surface $j_{z}$.  Summing 
over all contributions, Eq.  (\ref{areaop}) becomes \cite{QGI,FLR}
\begin{equation}
	\label{areaspec}
	\hat{A}_{S} \ket{s} 
	= \frac{l^{2}}{2} \sum_{v \in S \cap {\mathsf G}} 
	\left[ 2 j_{p_{v}}(j_{p_{v}} +1) + 2 j_{n_{v}}(j_{n_{v}} +1) -
	j_{z_{v}}(j_{z_{v}} +1) \right]^{1/2} \ket{s}.
\end{equation}
This result suggests that space is discrete; measurements of area can only
take quantized values.  Other geometric operators such as length 
\cite{Tlength}, angle \cite{SAMangle}, and volume share this 
property. 

The construction of the volume operator is more complicated than the 
area construction.  Nevertheless, it is possible, starting from the 
classical expression
\[
V_{R} =  \int_{R} d^{3}x \sqrt{q} = \int_{R} d^{3}x \sqrt{ \frac{1}{3!}
\epsilon_{abc} \epsilon^{ijk} E^{ai} E^{bj} E^{ck}},
\]
to regularize the quantity and define the operator \cite{RSareavol}, 
\cite{QGII} - \cite{BorissovetalMatrix}\footnote{Any differences in 
definition of the volume operators, such as those arising in the placement 
of the absolute value or in the regularization, do not play a critical 
role in the introduction of the quasilocal energy operator.  The 
Rovelli-Smolin form of the operator is given here \cite{RSareavol,RC}.}
\[
\hat{V}_{R} \ket{s} = (4 \pi G )^{\frac{3}{2}} 
\sum_{v \in {\mathsf G}} 
\sqrt{ \left| \frac{1}{8 \cdot 3!} \sum_{I,J,K \dashv v} 
\chi^{{}}_{IJK} \epsilon_{ijk} \hat{J}_{I}^{i} \hat{J}_{J}^{j} \hat{J}_{K}^{k}
\right|} \ket{s}
\]
where the sign factor is given by $\chi_{IJK} = {\rm 
sgn}(\det(\dot{e}_{I}(0), \dot{e}_{J}(0), \dot{e}_{K}(0))$. (Incident 
edges are oriented outward for ease of writing.)  The spectra 
of such an operator can be worked out using recoupling theory as in 
Refs.  \cite{RC}, \cite{BorissovetalMatrix}, and \cite{Tvol}.  
Introductions to diagrammatic recoupling theory may be found in 
Refs.  \cite{KL,SAMsn}.

\section{The quasilocal energy operator}
\label{qe}

To begin, I give a regularization of the Hamiltonian $H_{S} (N) $ with 
a scalar lapse.  This term may be made into a well-defined quantum 
operator on the appropriate state space.  However, it does not have 
the dimensions of energy.  Two normalizations to fix this problem are 
explored, one using area and the other using volume.  In both cases 
the full spectrum may be computed.  There are differences.  The 
operator which, in the naive classical limit, gives the familiar 
expression for the ADM energy and quasilocal energy requires a 
restriction on the Hilbert space.  The alternate operator is defined 
on the full Hilbert space but does not have the correct naive 
classical limit.  This is discussed in Sections \ref{qevol} and 
\ref{conclusion}.  The next section provides a regularization of 
$H_{S}(N)$.  A recoupling identity studied in \ref{recouplingnote} 
suggests the final form of the operators.  The spectra are given in 
Section \ref{spectra}.

\subsection{Regularization of the classical observable}
\label{reg}

The classical quantity to be promoted to a quantum operator is given 
by
\[
H_{S} (N) = \frac{1}{4 \pi G} 
\int_{S} d^{2}x N n_{a}(x) 
\epsilon^{ijk} \left( A_{b}^{i} \tilde{E}^{bj} \tilde{E}^{ak} \right)(x).
\]
(For the remainder of this paper I explicitly give the density weights 
using tildes.  This notation also serves to distinguish the two forms 
of the quasilocal energy operator.)  The regularization is based on the 
observation that the boundary Hamiltonian has two parts.  One part, 
$n_{a}\tilde{E}^{ak}$, is the familiar triad operator integrated over 
a 2-surface as in Eq.  (\ref{Edef}).  The second part, 
$A_{a}^{i}\tilde{E}^{aj}$, may be roughly described as a projection of 
the connection along an edge of a spin network.  These two parts are 
regularized by point splitting.

The regularization of the classical Hamiltonian is in several steps.  
The surface is thickened, as in the definition of the area operator in 
Ref.  \cite{FLR}.  This is done by introducing a smooth coordinate $r$ 
over a finite neighborhood of $S$ with ``thickness'' $\tau$.  The 
boundary itself is located at $r=0$.  The thickened surface has a 
natural foliation in terms of $r$ and gives a way to regulate the 
operator.  This portion of the regularization handles the tangent space 
factor (the sign $\chi_{S}$).  The boundary integration is transformed 
to an integration over the thickened surface, a region ${\cal R}$:
\[
\int_{S} d^{2} \sigma \rightarrow \frac{1}{\tau} 
\int_{-\tfrac{\tau}{2}}^{\tfrac{\tau}{2}} dr \, \int_{S} 
d^{2} \sigma \equiv \frac{1}{\tau} \int_{\cal R} d^{3}x.
\]
The region ${\cal R}$ is further partitioned into cells ${\cal R}_{c}$ 
which have the property that the coordinate width, denoted by 
$\epsilon$, is tied to the coordinate height so that $\tau = 
\epsilon^{k}$.  The power $k$ is restricted to lie between 1 and 
2.\footnote{As will be clear in the following discussion the reason 
for this restriction is identical to the one used in Ref.  
\cite{FLR}.} By construction, the number of cells $n_{\epsilon}$ is 
tied to the same limit.  The fully regularized operator is 
averaged over the leaves of the foliation of ${\cal R}$ and point split.
The classical expression is defined by
\begin{equation}
	\label{hdef}
	\left[ H_{S} (N) \right]_{\epsilon} 
	:= \frac{1}{4 \pi G} \frac{1}{\tau} 
	\sum_{c=1}^{n_{\epsilon}}
	\int \limits_{{\cal R}_{c} \otimes {\cal R}_{c}} \! \! \! \!
	d^{3}x \, d^{3}y \, 
	\epsilon^{ijk} (A_{a}^{i} \tilde{E}^{aj})(y) \, 
	N(x) \, n_{b}(x) \tilde{E}^{bk}(x).
\end{equation}
The regulated Hamiltonian is well-defined and goes to Eq.  
(\ref{hclass}) in the limit as $\epsilon$ vanishes.

In the next steps I re-express the classical variables $A_{a}^{i}$ and 
$\tilde{E}^{ai}$ in a suitable form for quantization.  Making use of 
the identity $\epsilon^{ijk}= (-2){\rm Tr} \left[ [\tau^{i}, \tau^{j}] 
\tau^{k}\right]$, $\left[ H_{S} (N) \right]_{\epsilon}$ is transformed 
into the convenient expression
\begin{equation*}
		\left[ H_{S} (N) \right]_{\epsilon} 
		= - \frac{1}{2 \pi G} \frac{1}{\tau}   
		\sum_{c=1}^{n_{\epsilon}} \,
		\int \limits_{{\cal R}_{c} \otimes {\cal R}_{c}} \! \! \!
		d^{3}x \, d^{3}y \, N(x) n_{b}(x)
		{\rm Tr}\left[ [A_{a},\tilde{E}^{a}](y) \tilde{E}^{b}(x) \right].
\end{equation*}
Using the triad integrated over a surface, Eq.  (\ref{Edef}), and the 
action on one edge, Eq.  (\ref{triadbracket}), a short calculation 
shows that
\begin{equation} 
	\label{step4}
	\begin{split}
	\left[ H_{S} (N) \right]_{\epsilon} =
	- \frac{1}{2 \pi G}(4 \pi G)^{2} \frac{1}{\tau} 
	\sum_{c=1}^{n_{\epsilon}} \,
	\int \limits_{e_{I} \in {\cal R}_{c} } \! \! \! dt 
	\int \limits_{e_{J} \in {\cal R}_{c} } \! \! \! ds 
	N(e_{I}(t)) n_{b}(e_{I}(t)) \, \dot{e}_{I}^{b}(t) \\ 
	\times {\rm Tr}\left[ [A_{a} (e_{J}(s)) \dot{e}_{J}^{a}(s), J_{J}] 
	J_{I} \right].
	\end{split}
\end{equation}
Next, the ``edge-projected connection'' $A_{a}\dot{e}_{J}^{a}$ is 
expressed in terms of holonomies.  This can be done in a 
straightforward manner as smooth connections satisfy
\[
U_{e}(s, s + \delta s)^{\pm 1} = 1 \pm \delta s \, \dot{e}^{a}(s) 
A_{a} (e(s)) + O(\delta s^{2}).
\]
The partitioning of the thickened surface splits the 
integration along the edges in Eq.  (\ref{step4}) into segments of 
maximum coordinate length $\epsilon$.  The integral $\int ds$ is 
approximated, in the limit, by $\sum \epsilon$.

Meanwhile, the integral along the edge $e_{I}$ in $t$ contains the 
geometrical factor
\begin{equation*}
	\begin{split}
		\int \limits_{e \in {\cal R}_{c}} dt \,
		n_{a}(e(t)) \, \dot{e}^{a}(t) 
		&=
		\begin{cases}
			+\tau & \text{when $\dot{e}^{a}$ is aligned with $n_{a}$}\\
			0 &  \text{when the edge is tangent}\\
			- \tau & \text{when $\dot{e}^{a}$ is anti-aligned with $n_{a}$}
		\end{cases} \\
		&\equiv 
		\chi^{S}_{e} \, \tau.
	\end{split}
\end{equation*}
This result is a consequence of the construction of the 
cells.\footnote{This is the reason for the restriction on the relation 
between $\tau$ and $\epsilon$ \cite{FLR}.} As is clear from this 
expression, the action of the regularized expression is, in the limit, 
only non-vanishing when an edge or a vertex of the cylindrical 
function's graph ${\mathsf G}$ contains a transverse intersection with 
the bounding surface.  For a finite graph, the sum over cells becomes 
a finite sum over these intersections.

As $\epsilon$ tends to zero $\left[ H_{S} (N) \right]_{\epsilon}$ of 
Eq.  (\ref{step4}) becomes\footnote{$J$ is or, shortly is promoted to, 
an operator.  This means that the leading order contribution of the 
factor $U_{e^{}_{J}}J_{J}U_{e_{J}^{-1}}$ is linear in $\epsilon$.  If 
desired one can include an explicit subtraction term $-J_{J}J_{I}$.  
The resulting operator is identical in action, but the spectrum differs 
from the one considered here by a factor of 2.}
\begin{equation}
	\label{Acomponent}
	- 8 \pi G \sum_{v \in S \cap {\mathsf G}} 
	\sum_{I,J \dashv v}
	N_{v} \chi_{I}^{S}
	{\rm Tr}\left[ U_{e^{}_{J}}J_{J}U_{e_{J}^{-1}}
	J_{I} \right] + O(\epsilon^{2})
\end{equation}
in which $N_{v}$ is the value of the lapse at the vertex $v$.  The 
quantum expression may then be defined as
\begin{equation}
	\label{quanthdef}
	\hat{H}_{S} (N) \ket{s} := 
	- 8 \pi G 
	\sum_{v \in S \cap {\mathsf G}} 
	\sum_{I,J \dashv v}
	N_{v} \chi_{I}^{S}
	{\rm Tr}\left[ U_{e^{}_{J}} \hat{J}_{J}U_{e_{J}^{-1}} 
	\hat{J}_{I} \right] \ket{s}
\end{equation}
on a spin network state $\ket{s}$.  The regularization in which the 
classical Hamiltonian is averaged over a thickened surface and 
point-split produces a gauge invariant, seemingly well-defined 
boundary Hamiltonian.  It is a new operator on the kinematic state 
space.  Its action has an intriguing ``self-measuring'' property: 
reading from left to right, the operator grasps an edge then alters 
the state by adding the edge, $e_{J}^{-1}$.  The operator measures the 
angular momentum of the edge which it modified.  Before exploring the 
details of the action, a more obvious problem must be addressed.

This operator does not have the correct dimension.\footnote{The 
quantum ``Hamiltonian'' has dimensions $\hbar G$ -- the $\hbar$ is in 
the definition of the operator $J$ -- so has the units of area rather 
than the dimensions $\sqrt{\hbar / G}$ of energy.} The source of this 
trouble is the (neglected) density weight tucked into the lapse; the 
expression is missing a factor of dimension inverse volume.  There are 
two ways to fix this problem.  First, as the energy is defined on a 
2-surface, the natural unit of length is provided by the area operator 
of Eq.  (\ref{areaop}) which may be used to normalize the surface 
term.  The quantum quasilocal energy operator on a surface $S$ is 
defined by
\begin{equation}
	\label{qeadef0}
	\hat{E}_{S}(N) \ket{s} := 
	- 8 \pi G 
	\sum_{v \in S \cap {\mathsf G}} \; 
	\sum_{I,J \dashv v}
	\frac{N_{v} \chi_{I}^{S}}{ \left[ \sqrt{\hat{A}_{v}} \right]^{3}}   
    {\rm Tr}\left[ U_{e_{J}}\hat{J}_{J}U_{e_{J}}^{-1}
    \hat{J}_{I} \right] \ket{s}.
\end{equation}
The operator $\hat{A}_{v}$ acts only on the vertex $v$.  As the area 
operator is a well-defined operator on the state space, it can be 
replaced with its spectral resolution.  The second solution to the 
correct normalization is to include the volume factor in the 
regularization.  A regularization of the latter form is given in 
Section \ref{qevol}.

\subsection{A recoupling identity}
\label{recouplingnote}

This section is devoted to a recoupling identity which provides a more 
tractable form of the quasilocal energy operator defined in the last 
section.  The definition simplifies dramatically.  The techniques 
employed are those of diagrammatic recoupling theory introduced 
to canonical quantum gravity in Ref.  \cite{BMSgqsn} and further 
developed in Refs.  \cite{BorissovetalMatrix,RC,roberto,SAMsn}.  A 
complete development of the techniques in terms of ``framed spin 
networks'' or Temperley-Lieb recoupling theory is in Ref. \cite{KL}.

Starting from the definition of the quasilocal energy operator Eq.  
(\ref{qeadef0}) a direct, but moderately lengthy, recoupling 
calculation yields the full spectrum.  However, there is a shorter 
method.  The idea is to use recoupling to investigate and re-express 
the action of
\[
{\rm Tr}\left[ U_{e_{J}} \hat{J}_{J}U_{e_{J}}^{-1} \hat{J}_{I} \right].
\]
This factor, acting on the edge $e_{J}$, can be reduced to a simple 
form.  To see this, note that the two hands, $\hat{J}_{I}$ and 
$\hat{J}_{J}$, grasp edges independently.  The unusual action is 
entirely contained in $U_{e_{J}}\hat{J}_{J}U_{(e_{J})^{-1}}$ which has 
the effect of overlaying a segment of the edge $e_{J}$, grasping the 
altered edge, and then retracing the segment.  The overlaying can 
easily be accounted for using the ``edge addition formula'' 
\cite{KL,BMSgqsn,dis}
\[
\kd{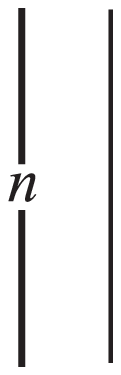} = \kd{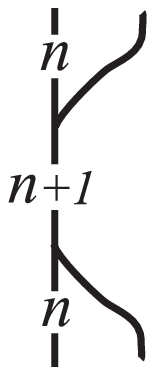} - \frac{n}{n+1} \kd{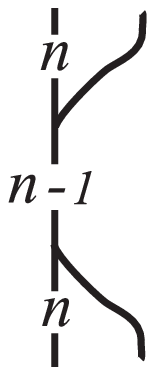}.
\]
Thus, when the holonomy $U_{(e_{J})^{-1}}$ acts on the edge $e_{J}$ 
(labeled by $n$), the result is the linear combination
\[
\kd{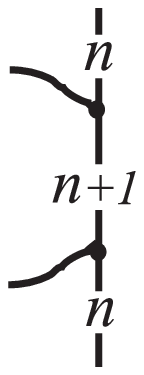} - \frac{n}{n+1} \kd{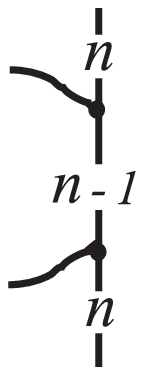}.
\]
The operator $\hat{J}_{J}$ acts next.  It is tied to the same edge and 
so measures the angular momentum on the edge that was 
modified.\footnote{While $\hat{J}_{J}$ must act on the edge $e_{J}$, 
there is some ambiguity in where, diagrammatically, it grasps the 
edge.  Not surprisingly, the results are the same, even when the 
operator grasps the edge above or below the modified edge.} The edge 
$e_{J}$ is grasped, yielding a 2-line and an overall factor of $n 
\pm 1$
\[
(n+1) \kd{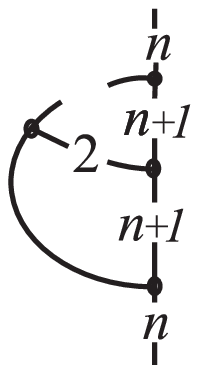} - n \, \frac{n-1}{n+1} \kd{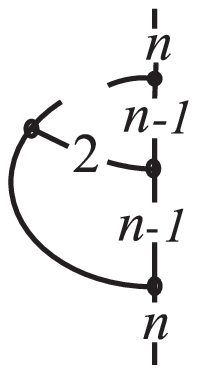}.
\]
Finally, the ``loop'' ($U_{e_{J}} \, U_{e_{J}^{-1}}$) is closed and
the second hand is included to give the result
\begin{equation}
	\label{recoupident1}
	(n+1) \kd{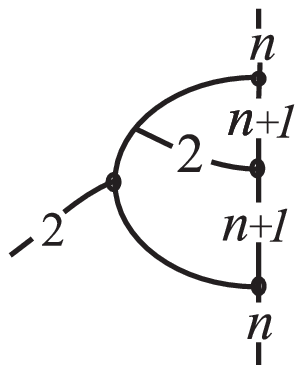} - n \, \frac{(n-1)}{(n+1)} \kd{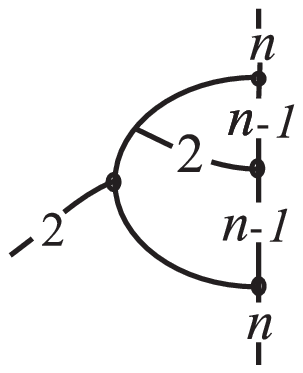}.
\end{equation}
This is the diagrammatic action of the factor $U_{e_{J}}
\hat{J}_{J}U_{e_{J}}^{-1}\, \hat{J}_{I}$.  By the Wigner-Eckart
theorem each of these diagrams is equivalent to a single trivalent
vertex \cite{SAMsn,dis,JM}.  Diagrammatically, the reduction is
accomplished in two stages, each of which removes a triangular loop. 
The first is given by
\[
\kd{si+} = - \frac{(n+3)}{2(n+1)} \kd{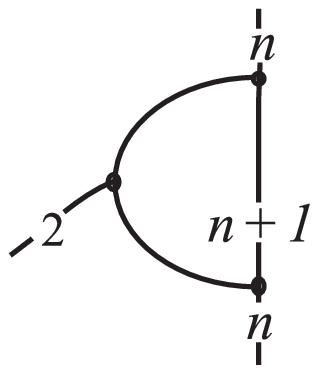}.
\]
The needed recoupling coefficients are given in the appendix (See Eqs. 
(\ref{tetident}) and (\ref{Teteg})).  Inserting these results in Eq. 
(\ref{recoupident1}), one finds
\begin{equation}
	\label{recoupleident}
	\begin{split}
		& (n+1) \kd{si+} - n\frac{(n-1)}{(n+1)} \kd{si-}\\
		&= \left[ - n \frac{(n+3)}{2(n+1)} - n \frac{(n-1)}{2(n+1)} \right]
		\skd{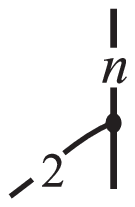} \\
		&= - n \skd{2hand}.
	\end{split}
\end{equation}
Therefore, we learn that the numerator of the quasilocal energy is 
most simply expressed in terms of a signed, ``spin-orbit coupling'' term:
\[
	{\rm Tr}\left[ U_{e_{J}} \hat{J}_{J}U_{e_{J}}^{-1} \hat{J}_{I} \right]
	\equiv 
	\frac{1}{2} \hat{J}^{i}_{I} \hat{J}^{i}_{J}
\]
in which I have used the identity $ {\rm Tr} \tau^{i} \tau^{j} = - 
\tfrac{1}{2} \delta^{ij}$.  The additional sign in this expression is 
due to the way traces are represented in the diagrammatic algebra:  
whenever a loop is closed, an addition sign is included (see, for 
instance, Ref. \cite{BorissovetalMatrix}).

This identity suggests the final form of the quasilocal energy 
operator
\begin{equation}
	\label{qeadef}
	\hat{E}_{S}(N) \equiv -  (4 \pi G) 
	\sum_{v \in S \cap \mathsf{G}} 
	\sum_{I,J \dashv v} N_{v}
	\chi^{S}_{I} \frac{\hat{J}_{J} \cdot \hat{J}_{I}}
	{\left(\sqrt{ \hat{A}_{v}}\right)^{3}}.
\end{equation}
The properties of the operator are explored in Section \ref{spectra} 
after the ``volume normalized'' operator is in place.  At this stage 
it is worth commenting on the operator ordering in Eq.  
(\ref{qeadef0}) which now can be phrased in terms of the recoupling 
identity.  The natural orderings of the expression are given by the 
cyclic permutations of the operators in the trace.  A short 
investigation shows that there is one ordering which is distinct from 
the action given in the last section: The order ${\rm Tr}[ 
U_{e^{-1}_{J}} \hat{J}_{I} U_{e_{J}} \hat{J}_{J}]$ differs in that, 
diagrammatically, $\hat{J}_{J}$ grasps the edge $e_{J}$ before the 
operator overlays an edge.  This produces a diagram distinct from 
those above.  Nevertheless, the two terms created with the edge addition 
formula trivially re-sum to give a loop in the 2-line of the operator.  
Such a loop in a 2-line is equivalent to a 2-line.  Therefore, this is 
again proportional to ${\rm Tr} [ \hat{J}_{J} \hat{J}_{I} ]$.

We have seen that the regularization of the boundary Hamiltonian with
scalar lapse led almost directly to this result, Eq.  (\ref{qeadef}). 
The key element of choice is the ``normalization.''  In the next
section I regulate the true boundary Hamiltonian.  As we will see,
this regularization is successful but the resulting operator requires
a strong restriction on the state space.

\subsection{Alternate regularization and operator}
\label{qevol}

There is an alternate regularization of the operator in which the 
density is no longer absorbed by the lapse.  In this case the 
classical quantity to be quantized is
\[
H_{S}(\Ntil \, )   
= \frac{1}{4 \pi G} \int_{S} d^{2}x \, N 
\epsilon^{ijk} n_{b} \frac{A_{a}^{i} \tilde{E}^{aj} 
\tilde{E}^{bk}}{\sqrt{q}}.
\]
The density factor is also promoted to an operator.  To regularize 
this expression one may follow Thiemann \cite{QSDVI} and 
simultaneously soften the divergence of $\sqrt{q}$ and point-split 
with
\[
V_{\epsilon}(x) := \int_{\Sigma} d^{3}y f_{\epsilon}(x,y) \sqrt{q}(y)
\]
in which the smoothed characteristic function $f_{\epsilon}$ satisfies 
\[
\lim_{\epsilon \rightarrow 0} \frac{f_{\epsilon}(x,y)}{\epsilon^{3}}
= \delta^{(3)}(x,y).
\]
Note that $ \lim_{\epsilon \rightarrow 0} V_{\epsilon}(x) / 
\epsilon^{3} =\sqrt{q}(x)$.  Proceeding in a manner similar to the 
regularization in Section \ref{reg}, the two parts are point-split to 
give the regularized quasilocal energy
\[
\left[ H_{S}(\Ntil \, )  \right]_{\epsilon} := \frac{-1}{2 \pi G}
\int_{S} d^{2}x \, n_{b}(x) \frac{N(x)}{\sqrt{q}(x)}
\int d^{3}y \frac{f_{\epsilon}(x,y)}{\epsilon^{3}} {\rm Tr}
\left[ [A_{a}, \tilde{E}^{a}](y) \tilde{E}^{b}(x) \right].
\]
To reach the quantum operator, the triads and connection are expressed 
in forms suitable for quantization.  Replacing the triads with the 
form in Eq.  (\ref{triadbracket}) and the factor $\epsilon^{3} 
\sqrt{q} (x)$ with $V_{\epsilon}(x)$, one finds that
\begin{equation*}
	\begin{split}
		\left[ H_{S}(\Ntil \, )  \right]_{\epsilon} &=
		- (8 \pi G) 
		\sum_{I,J}
		\int_{S} d^{2}x \, n_{b}(x) 
		\frac{N(x)}{V_{\epsilon}(x)} 
		\int_{e_{I}} \! \! ds \, 
		\dot{e}_{I}(s) \tilde{\delta}(x, e_{I}(s)) \\
		& \times \int d^{3} y  \int_{e_{J}} \! \! dt \, 
		\tilde{\delta}(y, e_{J}(t)) 
		f_{\epsilon}(x, y) {\rm Tr}
		\left[ [A_{b}(y) \dot{e}^{b}_{J}(t), J_{J} ] J_{I} \right] \\
		&= - (8 \pi G) 
		\sum_{I,J}
		\int_{S} d^{2} x \, n_{b}(x) 
		\frac{N(x)}{V_{\epsilon}(x)}\\
		& \times \int_{e_{I}} \! \! dt \, \dot{e}_{I}^{b}(t) 
		\tilde{\delta}(x, e_{I}(t) ) 
		\int_{e_{J}} \! \! ds \, f_{\epsilon}(x, e_{J}(s)) {\rm Tr}
		\left[ [A_{b}(e_{J}(s)) \dot{e}^{b}_{J}(s), J_{J} ] J_{I} \right]
	\end{split}
\end{equation*}
with the delta function eliminating the $y$ integration in the second 
step.

The connection may be replaced by a holonomy.  This is possible as the 
factors inside the commutator are all evaluated on the edge $e_{J}$.  
It is also convenient to partition the edge $e_{J}$ using the 
definition of holonomy
\begin{equation*}
	\begin{split}
		\left[ H_{S}(\Ntil \, )  \right]_{\epsilon} 
		&= \lim_{n \rightarrow \infty}
		-(8 \pi G) 
		\sum_{I,J}
		\int_{S} d^{2}x \, n_{b}(x) 
		\frac{N(x)}{V_{\epsilon}(x)}
		\int_{e_{I}} \! \! dt \, \dot{e}_{I}^{b}(t) \delta(x, e_{I}(t) ) \\
		& \times \sum_{k=1}^{n} f_{\epsilon}(x, e_{J}(s_{k-1}))
		{\rm Tr}\left[ U_{e_{J}}(s_{k-1}, s_{k}) J_{J} 
		U_{(e_{J})^{-1}}(s_{k-1}, s_{k}) J_{I} \right] \\
		&= \lim_{n \rightarrow \infty}
		- (8 \pi G) 
		\sum_{v \in S \cap {\mathsf G}}
		\sum_{I\dashv v, J}
		\chi_{I}^{S} \frac{N_{v}}{V_{\epsilon}(v)}
		\sum_{k=1}^{n} f_{\epsilon}(v, e_{J}(s_{k-1}))\\
		& \times {\rm Tr}\left[ U_{e_{J}}(t_{k-1}, t_{k}) J_{J} 
		U_{(e_{J})^{-1}}(t_{k-1}, t_{k}) J_{I} \right]
\end{split}
\end{equation*}
where the usual definition for $\chi^{S}_{I}$ [Eq.  (\ref{chi})] is 
used.  I denote the intersection between the edge $e_{I}$ and 
the surface $S$ by $v$.  The volume $V_{\epsilon}(v)$ acts on the 
vertex $v$.

In the limit as $\epsilon$ vanishes, $f_{\epsilon}(v, e_{J})$ goes to 
$1$ if, and only if, the edge $e_{J}$ is incident to $v$.  Thus, only 
one term in the sum over partitions of the edge $e_{J}$ survives and 
the sum over of $J$ is tied to the vertex $v$.  Also in the limit, the 
volume goes to $V(v)$, the volume at the vertex $v$.  As the regulator 
is removed,
\[
\lim_{\epsilon \rightarrow 0} \left[ H_{S}(\Ntil \, )  \right]_{\epsilon}
= - (8 \pi G) 
	\sum_{v \in S \cap {\mathsf G}}
	\sum_{I, J \dashv v}
	\chi_{I}^{S} \frac{N_{v}}{V(v)}
	{\rm Tr}\left[ U_{e_{J}} J_{J} 	U_{(e_{J})^{-1}} J_{I} \right]
\]
in which the holonomy is defined to start at the incident end of 
$e_{J}$ (the ``germ'' of Ref. \cite{QGII}).

The quantization is now immediate.  Let me pull the units out of the 
geometric operator so that
\[
l^{3} \hat{V}_{v} := \lim_{\epsilon \rightarrow 0} V_{\epsilon}(v).
\]
Using the recoupling identity of Eq.  (\ref{recoupleident}),
quantum boundary Hamiltonian is defined as
\begin{equation} 
	\label{boundaryham}
	\hat{H}_{S}(\Ntil \, )
	= \frac{-1}{(\sqrt{ \pi \hbar^{3} G}}
	\sum_{v} 
	\sum_{I,J \dashv v} \chi_{I}^{S} 
	N_{v} \frac{{\rm Tr}[ \hat{J}_{J} \hat{J}_{I}]}{\hat{V}_{v}}.
\end{equation}
The reciprocal of the volume operator is evaluated using its spectral 
resolution.

Though the regulation of the boundary Hamiltonian is possible, the 
result suffers from the same difficulty as the $\hat{E}_{ADM}$ 
operator has in Ref.  \cite{QSDVI}:  The signed angular momentum 
operator $\chi^{S}_{I} \hat{J}_{I}$ does not commute with the volume.  
To prevent the whole operator from diverging, one can restrict the 
state space so that the graph has no edges tangent to the surface.  
This is the tangle property of Ref.  \cite{QSDVI}.  With this 
modified state space, denoted by $\ket{s_{i}}$, the 
quasilocal energy becomes
\begin{equation}
	\label{qevdef}
	\hat{E}_{S}( \Ntil \,) \ket{s_{i}} = \frac{1}{ \sqrt{4 \pi 
	\hbar^{3} G}} \sum_{v} \sum_{I,J \dashv v} N_{v} \chi^{S}_{I} 
	\frac{\hat{J}_{I} \cdot \hat{J}_{J}}{\hat{V}_{v}} \ket{s_{i}}
\end{equation}
The spin network state $\ket{s_{i}}$ satisfies the tangle 
property on $S$.  This is nearly identical to the $\hat{E}_{ADM}$ 
operator of Ref.  \cite{QSDVI}.  Aside from an overall factor, this 
operator differs from $\hat{E}_{ADM}$ in the sign $\chi^{S}_{I}$ and 
the lapse $N_{v}$.  Indeed, under the same asymptotically flat 
conditions as in Ref.  \cite{QSDVI} (unit lapse, tangle property, and 
all edges outgoing), this is $\hat{E}_{ADM}$.  

A detailed comparison between $\hat{E}_{S}( \Ntil \,)$ and 
$\hat{E}_{ADM}$ is in Section \ref{conclusion}, but it is worth 
offering one observation here.  In the context of a quasilocal 
operator the tangle property seems to be overly restrictive.  Since 
the quasilocal quantity may be applied to any surface in $\Sigma$, 
restricting the domain of this operator, and thus the Hilbert space, 
implies that there are no surfaces with edges inside them.  Since 
every edge lies within some surface, the property effectively 
eliminates the entire state space.  The tangle property is too strong 
for the quasilocal energy operator.

\subsection{The spectrum of quasilocal energy}
\label{spectra}

The operators of Eqs.  (\ref{quanthdef}), (\ref{qeadef}), and
(\ref{qevdef}) are quite similar to the geometric operators of
spin-net gravity and can be treated with the same methods.  The
spectra may be computed using the recoupling methods of Refs. 
\cite{RC,SAMsn,BMSgqsn} or the operator methods of Refs. 
\cite{QGI,QGII,QGIII}.  I first give the spectrum of the numerator,
the operator ``$\chi^{S}_{I} \hat{J}_{I} \cdot \hat{J}_{J}$.''  For
completeness, the calculation is in the space of gauge non-invariant
spin networks, the ``extended spin networks'' of Ref.  \cite{QGII}.

It is convenient to order the edges of each vertex into categories
according to the geometric factor $\chi^{S}_{I}$.  For a vertex of
valence $d$, the $a$ edges for which $\chi^{S}$ is positive are
labeled $e_{1}$ to $e_{a}$.  The $b$ edges for which $\chi^{S}$
vanishes are labeled $e_{a+1}$ to $e_{a+b}$.  Finally, the $c$ edges
for which $\chi^{S}$ is negative are labeled $e_{a+b+1}$ to $e_{d}$. 
There is no restriction on the order of edges within these partitions. 
Diagrammatically, this ordering is equivalent to selecting three
intertwiner trees which grow from internal edges labeled by
$p$, $z$, and $n$, according to the value of $\chi^{S}$.  The angular
momentum operators are similarly partitioned so that $\hat{J}_{(p)} :=
\hat{J}_{(e_{1})} + \hat{J}_{(e_{2})} + \dots + \hat{J}_{(e_{a})}$ for
edges with $\chi^{S}=1$; $\hat{J}_{(z)} := \hat{J}_{(e_{a+1})} +
\dots + \hat{J}_{(e_{a+b})}$ for edges with $\chi^{S}=0$; and
$\hat{J}_{(n)} := \hat{J}_{(e_{a+b+1})} + \dots + \hat{J}_{(e_{d})}$
for edges with $\chi^{S}=-1$.  Using the methods from the quantum
mechanics of angular momentum, one finds
\begin{equation}
	\label{Jform}
	\begin{split}
		\sum_{I,J} \chi^{S}_{I} \hat{J}_{I} \cdot \hat{J}_{J}
		&\equiv \left( \hat{J}_{(p)} - \hat{J}_{(n)} \right) \cdot 
		\left(\hat{J}_{(p)} + \hat{J}_{(z)} +\hat{J}_{(n)} \right) \\
		&= \hat{J}_{(p)}^{2} - \hat{J}_{(n)}^{2} + 
		\hat{J}_{(p)} \cdot \hat{J}_{(z)} - \hat{J}_{(n)} \cdot \hat{J}_{(z)}.
	\end{split}
\end{equation}
Two properties of this operator are immediately clear.  First, the
operator simplifies when evaluated on gauge invariant spin networks. 
In fact, the spectrum collapses and contains only the value zero. 
This is obvious from Eq.  (\ref{Jform}) as this operator contains
\begin{equation}
	\label{qGauss}
	\hat{G}^{i} \equiv \sum_{J} \hat{J}_{J}^{i}
\end{equation}
which is the quantum version of the Gauss constraint, Eq. 
(\ref{gauss}).\footnote{The Gauss constraint, although it is expressed
in terms of $\hat{J}$ operators, is edge orientation invariant.  In
the current work, the invariance comes from the ``edge-projected
connection'' in Eq.  (\ref{step4}).  When the orientation of the edge
$J$ is reversed, both the operator $\hat{J}_{J}$ and the component
along the connection change sign.} Of course, one may simply compute
the result as well.  For general vertices a recoupling calculation of
the terms in Eq.  (\ref{Jform}) shows that the operator vanishes.  The
recoupling terms are of two forms.  The first is diagrammatically
\begin{equation}
	\label{Jsquared}
	\hat{J}_{(n)}^{2} \skd{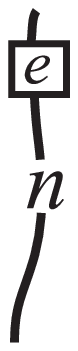} 
	= - \frac{\hbar^{2}}{2} n^{2} \skd{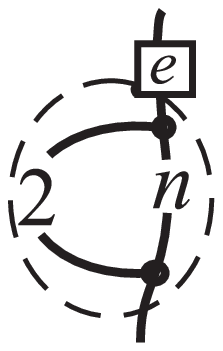}
	= \hbar^{2} \frac{n(n+2)}{4} \skd{lineen},
\end{equation}
making use of identity Eq. (\ref{bubnident}).  The second is
\begin{equation}
	\label{JdotJ}
	\begin{split}
	\hat{J}_{(p)} \cdot \hat{J}_{(z)} \skd{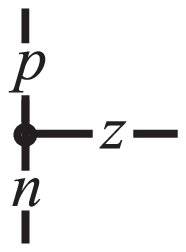} &= - \frac{\hbar^{2}}{2} 
	pz \skd{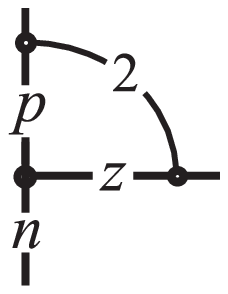} \\
	&= \frac{\hbar^{2}}{2} \frac{p(p+2) + z(z+2) - n(n+2)}{4} \skd{tripnz}
	\end{split}
\end{equation}
making use of identity Eq.  (\ref{trivident}).  Thus, as a consequence of
summing up all the components of the connection along the incident
edges (Eq.  (\ref{Acomponent})), the quasilocal energy operator
vanishes on gauge invariant states.

The second property seems equally serious.  Since the operator of Eq. 
(\ref{Jform}) contains two terms which do not commute, $\hat{J}_{(p)}
\cdot \hat{J}_{(z)}$ and $\hat{J}_{(n)} \cdot \hat{J}_{(z)}$, the
spectrum is not well-defined.  These terms cannot be simultaneously
diagonalized.  This result may be seen in the diagrammatic picture as
well.  In the gauge non-invariant case, vertices of this type have an
intertwiner with four edges, one of which is not grasped by either
operator.  (This extra edge can be thought of as leaving the
3-dimensional manifold.)  The operator of Eq.  (\ref{Jform}) requires
that all three edges labeled by $p$, $n$, and $z$ are part of a single
trivalent vertex.  This is not possible with a four-valent
intertwiner.  These properties suggest that the state space be 
carefully examined.

The space on which this operator acts is not characterized by general,
gauge invariant vertices.  When the spatial manifold is cut into
regions with boundary, the Hilbert space ${\cal H}^{I}$ of a bounded
region $I$ is based on ``open'' graphs (graphs with edges ending in
vertices of valence 1).  Given a graph in $\Sigma$, I define the open
subgraph ${\mathsf G}^{I}$ to be the portion of ${\mathsf G}$ within
the interior $I$ and within the boundary $\partial I$, i.e. ${\mathsf
G}^{I} = \overline{I} \cap {\mathsf G}$.  Thus, all the open edges are
incident to the boundary $S$ and these gauge non-invariant edges are
confined to the intersection of the graph with the boundary; the
``extended spin networks'' of Ref.  \cite{QGI} are restricted to lie
on the boundary.  Only edges that are incident to $v$ and in $G^{I}$
-- including tangent edges -- contribute to the energy.  The Hilbert
space is defined as before, only now the graph used is ${\mathsf
G}^{I}$.  With ${\cal H}^{I}$ it is possible to give the spectrum.

The clearest way to express spectrum of ``$\chi^{S}_{I} \hat{J}_{I}
\cdot \hat{J}_{J}$'' on ${\cal H}^{I}$ is to choose the orientations
of the edges incident to the boundary to be outward
pointing.\footnote{This is by no means necessary.  The spectrum can be
computed without assigning orientations although the computation is
considerably longer.} This is suggests a form for the intertwiner
``core.''\footnote{When the edges are partitioned into three
categories, as is often convenient in quantum geometry, the external
edges are connected in trees which end in one principle, internal
edge.  The core of the intertwiner is the trivalent vertex which
connects these three internal edges.  It is the only part of the
intertwiner which must be specified before completing the diagrammatic
calculation of the spectrum.} It is a trivalent vertex labeled by
$p,n$ and $z$ for the values of $\chi^{S}$.  With the orientations
outward pointing, then the labels $p$, $n$, and $z$ take the meanings
outside the surface, inside the surface, and tangent to the surface,
respectively.  As the $p$-edges are not in ${\cal H}^{I}$, they are
not seen by the operator.  Therefore the operator of Eq. 
(\ref{Jform}) reduces to
\begin{equation}
	\label{2Jform}
\hat{J}_{(n)} \cdot \hat{J}_{(p)}
\end{equation}
using $-\hat{J}_{(z)} = \hat{J}_{(n)} + \hat{J}_{(p)}$.  For
$\ket{s_{i}} \in {\cal H}^{I}$ based on a single vertex $v$ the
spectrum is
\begin{equation}
	\label{hispec}
	\sum_{I, J \dashv v}
	\chi^{S}_{I} \hat{J}_{I} \cdot \hat{J}_{J} \ket{s_{i}} 
	= \frac{\hbar^{2}}{2} \left[ \frac{p(p+2) + n(n+2) - z(z+2)}{4} \right]
	\ket{s_{i}}. 
\end{equation}
It is easy to generalize to more than one vertex; simply
sum over all contributing vertices.  The spectrum is not positive definite and,
acting on a single vertex, is explicitly bounded (for finite spin). 
There are two cases worth identifying.  When the vertex has no tangent
edges, the spectrum, which I call type (i), is always positive
definite and is proportional to $n(n+2)$.  The general case when there
are tangent edges, called type (ii), is given in Eq. 
(\ref{hispec}).

The form of the operator given in Eq.  (\ref{2Jform}) also provides a 
simple way to check two critical properties of the quasilocal energy, 
gauge invariance and compatibility with the area operator.  Gauge 
invariance is clear from Eq.  (\ref{Jform}), may be directly
checked,\footnote{The crux of the matter is that \[\left[ 
\hat{J}_{(p)} \cdot \hat{J}_{(n)}, \hat{G}^{i} \right] = - i \hbar 
\epsilon^{ijk} \left( \hat{J}_{(p)}^{j} \hat{J}_{(n)}^{k} + 
\hat{J}_{(p)}^{k} \hat{J}_{(n)}^{j} \right) =0.\]} and can 
also be made manifest with
\[
2 \hat{J}_{(p)} \cdot \hat{J}_{(n)}  =
\hat{J}^{2}_{(p)} + \hat{J}^{2}_{(n)} - \hat{J}^{2}_{(z)}.
\]
This expression is also useful to show that the two operators
$\chi^{S}_{I} \hat{J}_{I} \cdot \hat{J}_{J}$ and $\hat{A}_{v}$ are
compatible on the Hilbert space ${\cal H}^{I}$.  Thus, the terms in
the numerator and denominator may be diagonalized simultaneously.

It is now possible to assemble all the factors for the full spectrum 
of the quasilocal energy operator on ${\cal H}^{I}$.
For the operator defined in Eq. (\ref{qeadef}), using the spectrum of the 
area operator, Eq. (\ref{areaspec}), one has for type (i) 
\begin{equation}
	\begin {split}
	\label{qeaspec}
	\hat{E}_{S}(N) \ket{s} 
	&= m \sum_{v \in S \cap {\mathsf G}^{I}} 
	N_{v} \sqrt[4]{n_{v}(n_{v}+2)} \ket{s}. \\
\text{For type (ii):}\\
	\hat{E}_{S}(N) \ket{s}
	&=  m \sum_{v \in S \cap {\mathsf G}^{I}}
	N_{v}
	\frac{p_{v}(p_{v}+2) + n_{v}(n_{v}+2) - z_{v}(z_{v}+2)}
	{\left[ 2p_{v}(p_{v}+2) + 2n_{v}(n_{v}+2) - z_{v}(z_{v}+2) 
	\right]^{\frac{3}{4}}} \ket{s}. 
	\end{split}
\end{equation}
The fundamental mass scale is defined as
\[
m := \sqrt{ \frac{\hbar}{ 4 \pi G}}.
\]
The simple transverse vertices of type (i) yield a positive definite
spectrum while the others do not.\footnote{Note that the spectrum can
not be dependent only on the tangent edges; the operator vanishes
unless there is at least one transverse edge.} The results in Eq. 
(\ref{qeaspec}) are the full spectrum of the quasilocal energy
operator with area normalization.  The operator $\hat{E}_{s}(N)$ does
not require any restrictions on the Hilbert space.  It is well-defined
on the Hilbert space ${\cal H}^{I}$; all the needed properties of are
naturally induced from the full gauge invariant Hilbert space in
$\Sigma$.

There is one example of a spin-net state which is particularly 
interesting.  When the intersections between the graph of the spin 
network state and the surface are entirely transversal (type (i)), 
the quasilocal operator becomes
\begin{equation}
	\label{transversespec}
	\hat{E}_{S}(N) \ket{s_{i}} 
	=  m \sqrt{2} \sum_{v \in S \cap {\mathsf G}^{I}} 
	N_{v} \sqrt[4]{j_{v}(j_{v}+1)} \ket{s_{i}} \equiv
	m \sqrt{2} \sum_{v}  N_{v} \sqrt{a_{v}} \ket{s}
\end{equation}
in which $a_{v}$ is the eigenvalue of the area operator for the vertex
$v$.  The operator is the ``square root of the area''!  This could
have been anticipated with a little dimensional analysis. Nevertheless, it
is remarkable that energy is so closely related to area.  (The close
connection is also present in the volume normalization.)  For large
spins the energy scales as $\sqrt{j_{v} \,}$.

The next and final section begins with a summary of the operator 
definitions and continues with comparisons to earlier work and  
a discussion of some wider implications.

\section{Discussion}
\label{conclusion}

To be well-defined, the action of a theory must be functionally 
differentiable.  This simple observation provides a key to the form of 
all surface observables and boundary conditions of a given action 
\cite{HM,dis}.  In the case of the ($3+1$) gravitational theory 
defined in a bounded region, the variation generates a surface 
observable associated to the Hamiltonian constraint.  This term is 
precisely the negative of the Hamiltonian of the system.  In this 
paper, this boundary Hamiltonian is quantized within the framework of 
spin-net gravity.

Spin-net gravity is a background-metric independent, canonical 
quantization of gravity using real connection variables and the 
methods of spin networks.  Under this rubric is also included the 
assumption that the kinematic state space is rich enough to describe 
the full physical state space, including dynamics.

There are two different, inequivalent expressions for the boundary 
Hamiltonian corresponding to two separate normalizations.  Both 
operators are well-defined quantum operators (but on different 
spaces).  The area normalized operator has the following spectrum: For 
type (i) vertices:
\begin{equation*} 
	\hat{E}_{S}(N) \ket{s}
	=  m \sqrt{2 \gamma}
	\sum_{v} N_{v} \sqrt[4]{j_{v}(j_{v} +1)} \ket{s}
\end{equation*}
and for type (ii) vertices:	
\begin{equation*}
	\begin{split}
\hat{E}_{S}(N) \ket{s}
	= m \sqrt{2 \gamma} 
	\sum_{v} N_{v}
	[j_{(p_{v})}(j_{(p_{v})} +1) + j_{(n_{v})}(j_{(n_{v})} +1) 
	- j_{(z_{v})}(j_{(z_{v})} +1) ] \\
	\times \left[ 2 j_{(p_{v})}(j_{(p_{v})} +1) 
	+ 2 j_{(n_{v})}(j_{(n_{v})} +1)
	- j_{(z_{v})}(j_{(z_{v})} +1) \right]^{-3/4} \ket{s}.
	\end{split}
\end{equation*}
All the dimensionful constants and the Immirzi parameter are included;
$m = \sqrt{c \hbar/ 4 \pi G} \sim 40 GeV/c^{2}$.  This operator is
defined on the full Hilbert space ${\cal H}^{I}$.  This comes about
through an interplay between the operator and the gauge invariant
Hilbert space which allows the operator to be defined on the induced
Hilbert space.  As can be easily seen from the spectrum, only
transversal edges contribute energy.  Tangential edges remove energy.

The alternate form of the operator, with volume normalization and 
defined only on states which are of type (i), is
\begin{equation}
\hat{E}_{S}(\Ntil \,) \ket{s_{i}}= m \sqrt{2 \gamma} \sum_{v} N_{v}
	\frac{j_{(n_{v})}(j_{(n_{v})}+1)}{\lambda_{v}} \ket{s_{i}}	
\end{equation}
where $j_{(n_{v})}=n_{v}/2$ and $\lambda_{v}$ is the eigenvalue of the
volume.  On account of the state restriction, the numerator always has
the simple form of the area operator; the operator is proportional to
$\hat{A_{v}}^{2}/ \hat{V_{v}}$.  The volume operator does not have as
simple a form as the area operator, so the full spectrum cannot be
presented, as in Eq.  (\ref{qeaspec}).  The eigenvalues $\lambda_{v}$
can be computed using recoupling theory as in Refs.  \cite{RC,Tvol}. 
This form of the quasilocal energy has the correct naive classical
limit in that it is a direct quantization of the classical surface
Hamiltonian.  This operator requires tight restrictions on the Hilbert
space: There can be no edges tangent to the surface $S$.  As mentioned
in Section \ref{qevol}, since every edge is tangent to some surface,
this restriction introduces a contradiction in the 
construction of this quasilocal energy operator.

The operators share some general features.  They are defined on
bounding 2-surfaces.  Although they have a strange inclination to be
positive, both operators share the property that the spectrum is fully
discrete and bounded (for arbitrarily large but finite graphs and
spins).  The area normalized operator does not have a positive
spectrum.  The reason may be traced to the geometric factor in the new
operator $\chi^{S}_{I} \hat{J}_{I} \cdot \hat{J}_{J}$.  (This in turn
is a result of the Hamiltonian's two factors, the edge-projected
connection and the $\hat{E}^{i}_{S}$ piece.)  On account of this factor, the
quasilocal energy depends on the orientation of the surface.  It does
not, however, depend on the orientation of the edges.

The volume normalized operator has a similar form to the 
$\hat{E}_{ADM}$ energy of Ref.  \cite{QSDVI}, in which a different, 
but weakly equivalent, classical expression was quantized.  The 
present operator generalizes the ADM operator in one important way.  
It is a quasilocal operator defined on all bounding surfaces in the 
spatial manifold.  It also shares a key property: to be defined at all, 
the tangle property of Ref.  \cite{QSDVI} must be imposed.  In the 
language of Section \ref{spectra}, all vertices are of type (i).  In 
fact, when the tangle property is satisfied and when the lapse is 
fixed at unity, the expression for $\hat{E}_{S}(\Ntil)$ is simply 
identical to the $\hat{E}_{ADM}$ of Eq.  (3.18) in Ref.  \cite{QSDVI} 
(up to an overall numerical factor).  The volume normalized quasilocal 
energy reduces to the ADM energy.  It is remarkable that such 
different classical expressions yield nearly the same form of the 
operator.

Clearly, only one quasilocal energy operator gives physically 
correct values.  While $\hat{E}_{S}(\Ntil \,)$ has the correct naive 
limit, this criterion is not the only physical condition which must be 
met.  More importantly, the expectation value of the quantum energy 
operator in an appropriate semiclassical state must approximate the 
classical energy, up to small quantum corrections.  In addition, the 
algebra of boundary observables ought to be anomaly free.  Until such 
states and such operators are investigated in full detail, it is hard 
to definitively select a quasilocal energy operator.  To 
complete this investigation would require quantization of the boundary 
rotation and boosts -- a project which will be left to further work 
(there is preliminary work already in Ref.  \cite{QSDVI}).  Other 
approaches may also help fix the correct quantization, perhaps through 
the matching of quantum to semiclassical results.

In advance of a more complete investigation, there is one feature
which does differentiate the two operators.  The volume normalized
operator requires a restriction on the induced Hilbert space.  While
this may be acceptable for the asymptotically flat setting where the
property was introduced, this is a too severe {\it a priori}
restriction on the whole Hilbert space.  Therefore, for the remainder
of this section I restrict the majority of my comments to the area
normalized $\hat{E}_{S}(N)$.

The previous quantization of the boundary Hamiltonian has 
a qualitatively different action of ``shifting'' edges incident to the 
boundary \cite{BMP}.  The quasilocal energy operator does not share this 
qualitative behavior.  From the perspective of the spin-net framework, 
it is clear that the shift is an artifact of the partial quantization.  
In fact, there are two terms which shift ``up and back'' as may be 
seen in Eq.  (\ref{step4}) or in the gauge invariant form of 
the operator.  The ``shift'' is incorporated into the gauge invariant 
operator, which leads to the ``self-interaction'' measurement of edge 
addition as shown in Section \ref{recouplingnote}.

There are a few immediate results which follow from the 
definition of the quasilocal energy operator.  Gravitational energy 
takes on quantized values with the smallest gap in energy of
\[
e_{o} = \sqrt[4]{3} m
\]
due the the addition or removal of a spin-$1/2$ edge.  Suppose that a
system's bounding surface only intersected spin-$1/2$ edges of the
underlying spin network.  Then, transitions due to adding or removing a
spin-$1/2$ edge would result in radiation with the rather energetic
fundamental frequency of $\omega_{o} = \sqrt[4]{3} \, m / \hbar$. 
This mechanism is similar to the Bekenstein-Mukhanov quantization
\cite{BeMu}, in which the area is given by integer multiples of a
fundamental area.  In this case, however, the quasilocal energy is
quantized in integer units, $E_{S} = n e_{o}$, so area scales
as $n^{2}$.  Of course, before we tune our radio receivers to listen
to black holes, there is much more to understand.  In particular, we
need a characterization of semiclassical states and the dynamics.  To
see that this strongly affects the radiation, we need only study the
high spin limit \cite{highspin}.  The energy gap narrows, as would be
expected from the naive semiclassical limit.

There are also some wider implications of the quasilocal energy.  The
number of directions is enormous and I confine my comments to four
brief remarks: On a first glance at the quasilocal energy operator of
Eq.  (\ref{qeadef0}) it appears that the quasilocal energy operator
depends on the orientation of the edges.  Since the kinematic state
space of $SU(2)$ spin networks is independent of orientation this
would be odd.  In fact, the quasilocal energy operator does not depend
on orientation of the edges.  There are two aspects of this property. 
This is perhaps easiest to see by noting that when the orientation of
an edge is changed, both the sign factor $\chi_{I}^{S}$ and the
angular momentum operator $\hat{J}_{I}$ change sign \cite{QGII}.  In
an abuse of notation the same $\hat{J}_{I}$ are used in the Gauss
operator $\sum_{J} \hat{J}_{J}$, although the Gauss operator is
independent of edge orientation.  On the other hand, the operator
definitely does depend on the orientation of the surface.  When the
orientation of the surface changes, only the geometric factor changes
sign.  Orienting the edges to be outgoing from the vertices simply
places the edges in three, simple categories: inside the region,
outside the region, and tangent to the surface.  One tantalizing
aspect to this third category is the interpretation of edges which
leave the region $I$.  They take energy with them.  It remains to be
seen whether these could provide structure for the 4-dimensional space
-- ``pillars of time'' in a 4-dimensional spin network model -- or
more exotic manifolds.

Given the definition of the quasilocal energy operator and the induced
Hilbert space ${\cal H}^{I}$, the question arises whether the operator
is consistent.  If the spatial manifold $\Sigma$ is compact, is it
true that $E_{S} + E_{S^{*}}$ vanishes?  The natural language in which
to consider such questions is topological quantum field theory (TQFT)
\cite{TQFT}, in which one studies diffeomorphism invariant theories on
manifolds with boundary.  While the manifolds are cut and sewn
together, the theories associate maps to interiors and vector spaces
to boundaries.  A more careful study will be left to further work. 
But the question may addressed directly.  On a particular spin network
state, say $\ket{s_{\Sigma}}$, the two operators act on the two
``halves'' of the state.  One is based on $\overline{I}$; the other is
based on $\overline{I}^{*}$.  As $\ket{s_{\Sigma}}$ is a general,
gauge invariant state, there are two types of vertices to consider,
types (i) and (ii).  A short calculation shows that the contributions
to the energies are numerically identical.\footnote{Naturally, one
needs a consistent orientation on $\Sigma$ to perform the sum.}

As this quasilocal energy is the boundary Hamiltonian, one may 
describe the time development of the Lorentzian theory on $S \times 
{\mathbb R}$ 
\[
\hat{U}(t) = e^{i \hat{E}_{S} (N_{t}) t /\hbar }.
\]
(Recall that the boundary conditions fix the lapse and metric on the 
boundary.  A more complete treatment would also include terms with 
non-vanishing shift $N^{a}$, giving rotations and boosts.)  The energy 
operator is diagonal in the spin network basis, so the unitary 
evolution operator has a simple, well-defined form (at least for type 
(i) vertices).  It describes the system in terms of an observer on the 
surface $S$.

Likewise, the quasilocal energy operator gives the partition function 
$e^{\beta \hat{E}_{S}}$.  As the quasilocal energy is the true 
Hamiltonian, this function is the partition function for the 
statistical mechanics of spin-net gravity.  This observation is the 
starting point for a vast range of physical questions.  For instance, 
is the energy quanta $e_{o}$ statistically favored?  What is the 
selection criteria for the ground state(s)?  What is the entropy of a 
bounded system in spin-net gravity?

What is particularly striking is the delightful number of physical 
questions that can be addressed with this operator in the current 
framework on spin-net gravity.  This is but one result of the techniques 
which have been developed since 1995 and which offer methods for detailed 
study of these questions.  In fact, it appears that these techniques 
are powerful and rich enough to bring us from theoretical modeling to 
physical predictions.

\begin{ack} 
I thank Herbert Balasin and Robert DePietri for discussions.  Thanks 
also to an anonymous reviewer for helpful, critical remarks.  I 
gratefully acknowledge the support of the Austrian Science Foundation 
(FWF) through a Lise Meitner Fellowship.
\end{ack}

\appendix
\section{Recoupling quantities}

This appendix contains recoupling formulae needed in the 
calculations of the spectrum.  The conventions are those of
Kauffman and Lins \cite{KL} (for $A=-1$). Short introductions and 
definitions of the basic recoupling quantities can be found in 
\cite{RC,SAMsn}.

The diagrammatics for the ``$\chi \hat{J} \cdot \hat{J}$'' operator
require careful use of the ``$\lambda$-move''
\begin{equation} 
	\begin{split} 
		\label{lmove}
		\skd{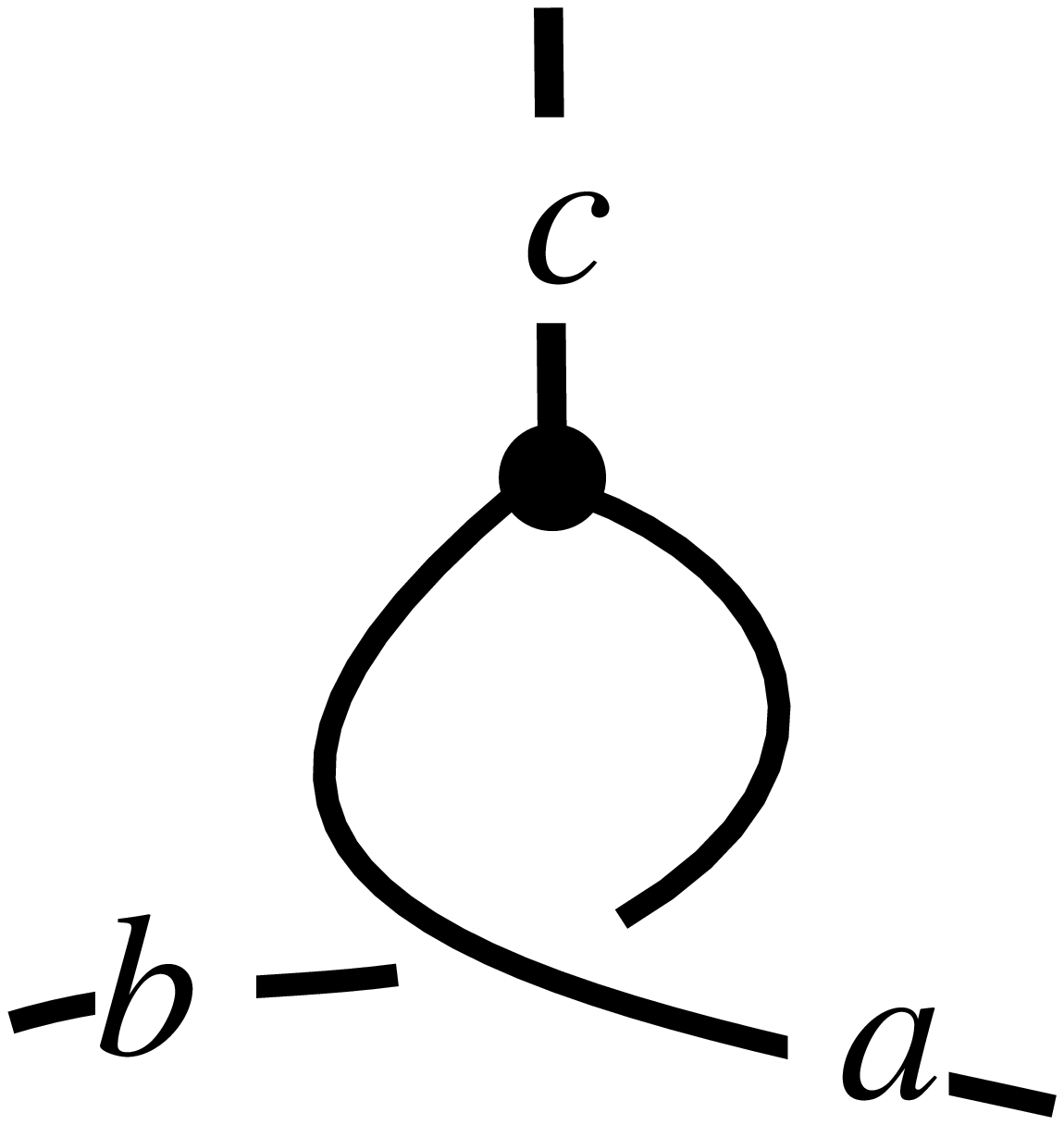} &= \lambda^{ab}_c \skd{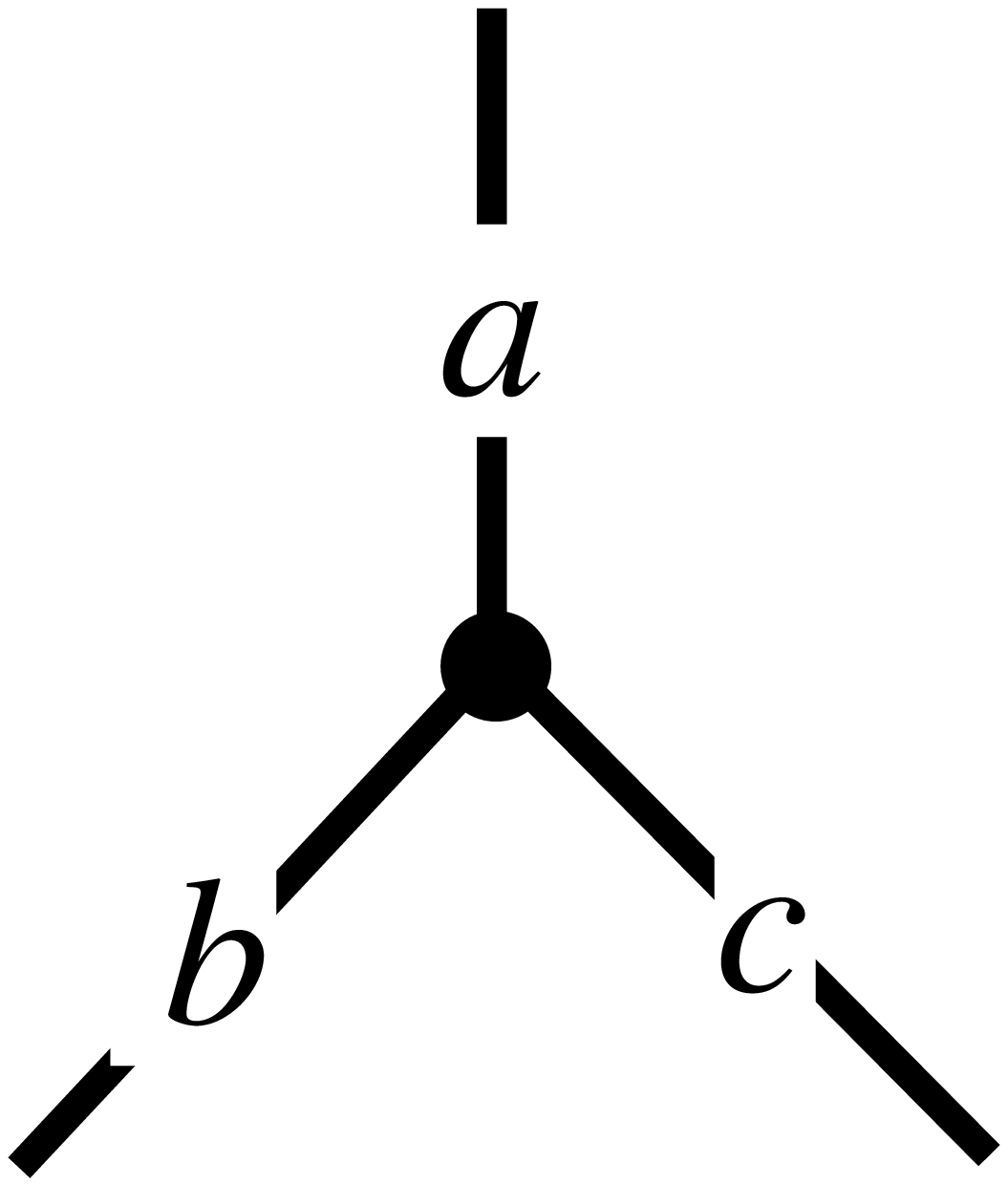}  
		\text{where $\lambda^{ab}_c$ is} \\
		\lambda^{ab}_c &= (-1)^{[a(a+3) + b(b+3) - c(c+3)]/2}.
	\end{split}
\end{equation}

The function $\theta(m, n, l)$ is given by
\begin{equation} \label{theta}
\theta(m,n,l)= 
(-1)^{(a+b+c)}{(a+b+c+1)!a!b!c! \over (a+b)!(b+c)!
(a+c)!}
\end{equation}
where $a=(l+m-n)/2$, $b=(m+n-l)/2$, and $c=(n+l-m)/2$.

The tetrahedral symbol is given by
\begin{equation} \begin{split}
\label{TetDef}
{\rm Tet} \begin{bmatrix} a & b & e \\ c & d & f \end{bmatrix} &= N 
\sum_{m \leq s \leq S} (-1)^s  {  (s+1)! \over
\prod_i\, (s-a_i)! \; \prod_j \, (b_j -s)! } \\
N &= { \prod_{i,j}\, [b_j - a_i]! \over a!b!c!d!e!f!}
\end{split} \end{equation}
in which
\begin{equation} \begin{align}
a_1 &= \tfrac{1}{2} ( a +d + e) & b_1 &= \tfrac{1}{2} ( b +d + e+ f) 
\nonumber \\
a_2 &= \tfrac{1}{2} ( b +c + e) & b_2 &= \tfrac{1}{2} ( a +c + e +f) 
\nonumber \\
a_3 &= \tfrac{1}{2} ( a +b + f) & b_3 &= \tfrac{1}{2} ( a +b + c+d) 
\nonumber \\
a_4 &= \tfrac{1}{2} ( c +d + f) & m={\rm max}\, \{a_i\} \ \ 
M={\rm min}\, \{b_j\} .\nonumber
\end{align} \end{equation}

The quantities needed in the calculation for the simplification of 
the quasilocal energy operator are:
\begin{equation}
	\label{tetident}
	\begin{split}
		\frac{{\rm Tet} 
		\begin{bmatrix} 
			1 & 1 & n \\ n+1 & n+1 & 2 
		\end{bmatrix}}
		{\theta(n+1,n,1)}
		&= - \frac{(n+3)}{2(n+1)};
		\; 
		\frac{{\rm Tet} 
		\begin{bmatrix}
			n & n & n+1 \\ 1 & 1 & 2 
		\end{bmatrix}}
		{\theta(n,n,2)}
		= \frac{n}{n+1} \\
		\frac{{\rm Tet} 
		\begin{bmatrix} 
			1 & 1 & n \\ n-1 & n-1 & 2 
		\end{bmatrix}}
		{\theta(n,n,2)}
		&= \frac{1}{2};
		\; \; \; \;
		\frac{{\rm Tet} 
		\begin{bmatrix}
			n & n & n-1 \\ 1 & 1 & 2 
		\end{bmatrix}}
		{\theta(n,n,2)}
		= 1.		
	\end{split}
\end{equation}
For instance,
\begin{equation}
	\label{Teteg}
	\begin{split}
\kd{si+} &= \frac{{\rm Tet} 
		\begin{bmatrix} 
			1 & 1 & n \\ n+1 & n+1 & 2 
		\end{bmatrix}}
		{\theta(n+1,n,1)} \kd{sitet}\\
		&= \frac{{\rm Tet} 
		\begin{bmatrix} 
			1 & 1 & n \\ n+1 & n+1 & 2 
		\end{bmatrix}}
		{\theta(n+1,n,1)}
		\frac{{\rm Tet} 
		\begin{bmatrix}
			n & n & n+1 \\ 1 & 1 & 2 
		\end{bmatrix}}
		{\theta(n,n,2)} \skd{2hand}.
	\end{split}
\end{equation}
This is used in the recoupling calculation of Eq. (\ref{recoupleident}).

A ``bubble'' diagram is proportional to a single edge.  In particular,
\begin{equation}
	\label{bubnident}
		\skd{opJJ} = \frac{\theta(n,n,2)}{\Delta_{n}} \skd{lineen}
		= - \frac{n+2}{2n} \skd{lineen}.
\end{equation}

Such a 2-line spanning a vertex is
\begin{equation}
	\label{trivident}
	\kd{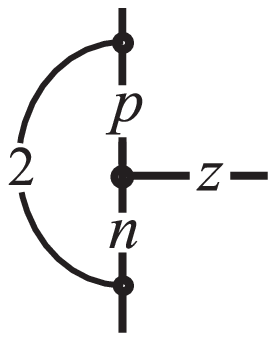} \equiv \frac{{\rm Tet} 
		\begin{bmatrix}
			p & p & z \\ n & n & 2 
		\end{bmatrix}}
		{\theta(p,n,2)} \skd{tripnz}
		= - \frac{ n(n+2) + p(p+2) - z(z+2)}{4 np} \skd{tripnz}.
\end{equation}


\begin{thebibliography}{999999}

\bibitem{bondi} H. Bondi, M.G.J. van der Burg, and A.W.K. Metzner,
{\em Proc. R. Soc. London} {\bf A269} (1962) 21.

\bibitem{sachs} R. K. Sachs, 
{\em Proc. R. Soc. London} {\bf A270} (1962) 103.

\bibitem{ADM} R. Arnowitt, S. Deser, and C. W. Misner, in 
{\em Gravitation, an Introduction to Current Research} edited by
L. Witten (Wiley, New York, 1962).

\bibitem{penroseQE} R. Penrose, ``Quasi-local mass and angular momentum in 
general relativity'' 
{\em Proc. Roy. Soc. London } {\bf A381} (1982) 53-63.

\bibitem{hawking} S. W. Hawking, {\em J. Math. Phys.} {\bf 9} (1968) 598-604.

\bibitem{geroch} R. P. Geroch, 
{\em Ann. N.Y. Acad. Sci.} {\bf 224} (1973) 108.

\bibitem{KO} J. Katz {\em Class. Quant. Grav.} {\bf 2} (1985) 423-425;
J. Katz and A. Ori {\em Class. Quant. Grav.} {\bf 7} (1990) 787-802.

\bibitem{JK} J. Jezierski and J. Kijowski {\em Gen. Rel. Grav.}
{\bf 22} 1283-1307.

\bibitem{DM} A. J. Dougan and L. J. Mason, ``Quasilocal mass 
constructions with positive energy'' {\em Phys. Rev. Lett.}
{\bf 67} 2119-2122.

\bibitem{B} G. Bergqvist, ``Positivity and definitions of mass'' 
{\em Class. Quant. Grav.} {\bf 9} (1992) 1917-1922.

\bibitem{NTJ} J. M. Nester and R. S. Tung {\em Gen. Rel. Grav.} {\bf 
27} (1995) 115-119; T. Jacobson and R. S. Tung {\em Class. Quant. Grav.} 
{\bf 12} (1995) L51-L55.

\bibitem{R} D. C. Robinson {\em Class. Quant. Grav.} {\bf 13} (1995) 
307-315.

\bibitem{hayward} S. A. Hayward, ``Quasilocal gravitational energy'' 
{\em Phys. Rev.} {\bf D 49} (1994) 831-839.

\bibitem{BY} J. D. Brown and J. W. York, ``Quasi-local energy and 
conserved charges derived from the gravitational action''
{\em Phys. Rev.} {\bf D 47}  (1993) 1407-1419.

\bibitem{lau} S. Lau {\em Class. Quant. Grav.} {\bf 10} (1993) 2379-2399.

\bibitem{HM} V. Husain and S. Major, 
``Gravity and BF theory defined in bounded regions''
{\em Nuc. Phys.} {\bf B 500} (1997) 381-401;
Online Preprint Archive: {\em http://xxx.lanl.gov/abs/gr-qc/9703043}.	

\bibitem{RT} T. Regge and C. Teitelboim, {\em Ann. Phys.} {\bf 88} 
(1974) 286.

\bibitem{NV} A. Ashtekar, {\it New perspectives in canonical gravity}, 
(Bibliopolis, Naples, 1988); { \it Lectures on non-perturbative canonical 
gravity} (World Scientific, Singapore, 1991).

\bibitem{barbero} F. Barbero, ``Real Ashtekar Variables for Lorentzian 
Signature Spacetimes'' 
{\em Phys. Rev.} {\bf D 51} (1995) 5498;
Online Preprint Archive: {\em http://xxx.lanl.gov/abs/gr-qc/9410014}.

\bibitem{CRrev} Carlo Rovelli, ``Loop Quantum Gravity,''
	{\em Living Reviews in Relativity} at
	{\em http://www.livingreviews.org/Articles/Volume1/1998-1rovelli};
	``Strings, Loops, and Others: A critical survey of the present 
	approaches to quantum gravity,'' in {\em Gravitation and Relativity: 
	At the turn of the Millennium}, Proceedings of the GR-15 Conference, 
	Naresh Dadhich and Jayant Narlikar, ed. (Inter-University Center 
	for Astronomy and Astrophysics, Pune, India, 1998), pp. 281 - 331,
	Online Preprint Archive: {\em http://xxx.lanl.gov/abs/gr-qc/9803024}.

\bibitem{Tlength} T. Thiemann, ``A length operator for canonical 
quantum gravity''
{\em J. Math. Phys.} {\bf 39} (1998) 3372-3392;
Online Preprint Archive: {\em http://xxx.lanl.gov/abs/gr-qc/9606092}. 

\bibitem{RSareavol} C. Rovelli and L. Smolin, ``Discreteness of area 
and volume in quantum gravity'' {\em Nuc. Phys.} {\bf B 422} (1995) 593;
Erratum {\em Nuc. Phys.} {\bf B 456} (1995) 753;
Online Preprint Archive: {\em http://xxx.lanl.gov/abs/qr-qc/9411005}.

\bibitem{QGI} A. Ashtekar and J. Lewandowski, 
``Quantum theory of geometry I: Area operators''
{\em Class. Quant. Grav.} {\bf 14} (1997) A43-A53;
Online Preprint Archive: {\em http://xxx.lanl.gov/abs/gr-qc/9602046}.

\bibitem{FLR} S. Frittelli, L. Lehner, C. Rovelli, ``The complete 
spectrum of the area from recoupling theory in loop quantum 
gravity'' {\em Class. Quant. Grav}. {\bf 13} (1996) 2921-2932;
Online Preprint Archive: {\em http://xxx.lanl.gov/abs/gr-qc/9608043}.

\bibitem{lollvol} R. Loll, ``The volume operator in discretized quantum 
gravity''
{\em Phys. Rev. Lett.} {\bf 75} (1995) 3048-3051;
Online Preprint Archive: {\em http://xxx.lanl.gov/abs/gr-qc/9506014} 
and ``Spectrum of the Volume Operator in Quantum Gravity''
{\em Nuc. Phys.} {\bf B 460} (1996) 143;
Online Preprint Archive: {\em http://xxx.lanl.gov/abs/gr-qc/9511030}. 

\bibitem{Lvol} Jerzy Lewandowski, ``Volume and Quantizations''
{\em Class. Quant. Grav.} {\bf 14} (1997) 71-76;
Online Preprint Archive: {\em http://xxx.lanl.gov/abs/gr-qc/9602035}.

\bibitem{Tvol} T. Thiemann, ``Closed formula for the matrix elements of 
the volume operator in canonical quantum gravity''
{\em J. Math. Phys.} {\bf 39} (1998) 3347-3371;
Online Preprint Archive: {\em http://xxx.lanl.gov/abs/gr-qc/9606091}.

\bibitem{BorissovetalMatrix} R. Borissov, R. De Pietri, C, Rovelli,
``Matrix Elements of Thiemann's Hamiltonian Constraint in Loop
Quantum Gravity''
{\em Class. Quant. Grav.} {\bf 14} (1997) 2793-2823;
Online Preprint Archive: {\em http://xxx.lanl.gov/abs/gr-qc/9703090}.

\bibitem{QGII} A. Ashtekar and J. Lewandowski, ``Quantum theory of 
geometry II: Volume operators'' {\em Adv. Theor. Math. Phys.} {\bf 1} 
(1998) 388;
Online Preprint Archive: {\em http://xxx.lanl.gov/abs/gr-qc/9711031}.

\bibitem{SAMangle} S. Major, ``Operators for quantized directions''
{\em Class. Quant. Grav.} in press;
Online Preprint Archive: {\em http://xxx.lanl.gov/abs/gr-qc/9905019}.

\bibitem{BMP}
John Baez, Javier P. Muniain, and Dardo Piriz, 
``Quantum Gravity Hamiltonian for Manifolds with Boundary''
{\em Phys. Rev} {\bf D 52} (1995) 6840-6845;
Online Preprint Archive: {\em http://xxx.lanl.gov/abs/gr-qc/9501016}.

\bibitem{QSDI} T. Thiemann, ``Quantum Spin Dynamics (QSD)''
{\em Class. Quant. Grav.} {\bf 15} (1998) 839-873;
Online Preprint Archive: {\em http://xxx.lanl.gov/abs/gr-qc/9606089}

\bibitem{QSDII} T. Thiemann, ``Quantum Spin Dynamics (QSD) II''
{\em Class. Quant. Grav.} {\bf 15} (1998) 875-905;
Online Preprint Archive: {\em http://xxx.lanl.gov/abs/gr-qc/9606090}

\bibitem{QSDIII} T. Thiemann, ``QSD III : Quantum Constraint Algebra 
and Physical Scalar Product in Quantum General Relativity''
{\em Class. Quant. Grav.} {\bf 15} (1998) 1207-1247;
Online Preprint Archive: {\em http://xxx.lanl.gov/abs/gr-qc/9705017}

\bibitem{QSDIV} T. Thiemann, ``QSD IV : 2+1 Euclidean Quantum Gravity 
as a model to test 3+1 Lorentzian Quantum Gravity''
{\em Class. Quant. Grav.} {\bf 15} (1998) 1249-1280;
Online Preprint Archive: {\em http://xxx.lanl.gov/abs/gr-qc/9705018}

\bibitem{QSDV} T. Thiemann, ``QSD V : Quantum Gravity as the Natural 
Regulator of Matter Quantum Field Theories''
{\em Class. Quant. Grav.} {\bf 15} (1998) 1281-1314;
Online Preprint Archive: {\em http://xxx.lanl.gov/abs/gr-qc/9705019}

\bibitem{QSDVI} T. Thiemann, ``Quantum Spin Dynamics (QSD): VI. 
Quantum Poicar\'e algebra and a quantum positivity of energy theorem 
for canonical quantum gravity''
{\em Class. Quant. Grav.} {\bf 15} (1998) 1463-1485;
Online Preprint Archive: {\em http://xxx.lanl.gov/abs/gr-qc/9705020}

\bibitem{Tasym} T. Thiemann, ``Generalized boundary conditions for 
general relativity for the asymptotically flar case in terms of 
Ashtekar's variables'' {\em Class. Quant. Grav.} {\bf 12} (1995) 
181-198.

\bibitem{weave} A. Ashtekar, C. Rovelli, and L. Smolin, ``Weaving a 
classical geometry with quantum threads'' {\em Phys. Rev. 
Lett.} {\bf 69} (1992) 237.

\bibitem{BM} H. Balasin and S. Major,
preprint in preparation.

\bibitem{QGIII} A. Ashtekar, A. Corichi, and J. A. Zapata, 
``Quantum Geometry III: Non-commutativity of Riemannian Structures'' 
{\em Class. Quant. Grav.} {\bf 15} (1998) 2955-2972;
Online Preprint Archive: {\em http://xxx.lanl.gov/abs/gr-qc/9806041}.

\bibitem{RC} Roberto DePietri and Carlo Rovelli, ``Geometry
eigenvalues and the scalar product from recoupling theory in loop
quantum gravity'' {\em Phys. Rev.} {\bf D 54} (1996) 2664-2690.
Online Preprint Archive]: {\em http://xxx.lanl.gov/abs/gr-qc/9602023}.

\bibitem{roberto}  R. DePietri, ``On the relation between the 
connection and the loop representation of quantum gravity,''
{\em Class. Quant. Grav.} {\bf 14} (1997) 53-70;
Online Preprint Archive: {\em http://xxx.lanl.gov/abs/gr-qc/9605064}. 

\bibitem{SAMsn} Seth A. Major, ``A Spin Network Primer''
{\em Am. J. Phys.} {\bf 67} (1999) 972-980;
Online Preprint Archive: {\em http://xxx.lanl.gov/abs/gr-qc/9905020}.

\bibitem{dis} S. A. Major, {\em ``q-Quantum Gravity''}, Ph.D. Dissertation,
The Pennsylvania State University (1997).

\bibitem{AI} A. Ashtekar and C. Isham, ``Representations of the holonomy 
algebras of gravity and non-abelian gauge theories'' {\em Class.
Quant. Grav.} {\bf 9} (1992) 1433-1485.

\bibitem{ALrep} A. Ashtekar and J. Lewandowski, ``Representation theory
of analytic holonomy $C^{\star}$ algebras'' in {\em Knots and quantum
gravity} J. Baez, ed. (Oxford University Press, 1994);
Online Preprint Archive: {\em http://xxx.lanl.gov/abs/gr-qc/9311010}.

\bibitem{baez} J. Baez, ``Generalized measures in gauge theory'' {\em Lett.
Math. Phys.} {\bf 31} (1994) 213-223.

\bibitem{ALproject1} A. Ashtekar and J. Lewandowski, 
``Projective techniques and functional integration for gauge theories''
{\em J. Math. Phys.} {\bf 36} (1995) 2170-2191.
Online Preprint Archive: {\em http://xxx.lanl.gov/abs/hep-th/9411046}.

\bibitem{ALProj}   A. Ashtekar and J. Lewandowski, ``Differential Geometry
on the Space of Connections via Graphs and Projective Limits'' 
{\em J. Geom. Phys.} {\bf 17} (1995) 191-230;
Online Preprint Archive: {\em http://xxx.lanl.gov/abs/hep-th/9412073}.   

\bibitem{ALMMT} A. Ashtekar, J. Lewandowski, D. Marolf, J. Mour\~ao,
T. Thiemann, ``Quantization of diffeomorphism invariant theories of 
connections with local degrees of freedom'' 
{\em J. Math. Phys.} {\bf 36} (1995) 6456;
Online Preprint Archive: {\em http://xxx.lanl.gov/abs/gr-qc/9504018}.

\bibitem{MM} J. Mour\~ao and D. Marolf, ``On the support of the Ashtekar-
Lewandowski measure'' {\em Comm. Math. Phys.} {\bf 170} (1995) 583.

\bibitem{BS} J. Baez and S. Sawin, ``Functional Integration for Spaces of
Connections;'' 
Online Preprint Archive: {\em http://xxx.lanl.gov/abs/q-alg/9507023}.

\bibitem{JS} T. Jacobson and L. Smolin, ``Nonperturbative quantum geometries''
{\em Nuc. Phys.} {\bf B 299} (1988) 295-345.

\bibitem{RSloop} C. Rovelli and L. Smolin, ``Loop representation of 
quantum general relativity'' {\em Nuc. Phys.} {\bf B 331} (1990) 80-152.

\bibitem{I} G. Immirzi, ``Quantum Gravity and Regge Calculus'' 
{\em Nuc. Phys. Proc. Suppl.} {\bf 57} (1997) 65-72.

\bibitem{RTI} C. Rovelli and T. Thiemann, ``The Immirzi parameter in quantum
general relativity''  {\em Phys.Rev.} {\bf D 57} (1998) 1009-1014.

\bibitem{GOPI} R. Gambini, O. Obregon, and J. Pullin, 
``Yang-Mills analogues of the Immirizi ambiguity''
{\em Phys. Rev.} {\bf D 59} (1999) 047505.

\bibitem{BMSgqsn} R. Borissov, S. Major and L. Smolin,
``The Geometry of Quantum Spin Networks''
{\em Class. Quant. Grav.} {\bf 13} (1996) 3183-3196;
Online Preprint Archive: {\em http://xxx.lanl.gov/abs/gr-qc/9512043}.

\bibitem{KL} Louis H. Kauffman and S\'ostenes L. Lins, 
{\em Temperley-Lieb Recoupling Theory and Invariants of 3-Manifolds},
Annals of Mathematics Studies N. 134 
(Princeton University Press, Princeton, 1994).

\bibitem{JM} John P. Moussouris, ``Quantum models as spacetime based 
	on recoupling theory'' Oxford DPhil Thesis, unpublished (1983).

\bibitem{BeMu} J. Bekenstein and V. F. Mukanhov,
{\em Phys. Lett.} {\bf B 360} (1995) 7.

\bibitem{highspin} M. Barreira, M. Carfora, and C. Rovelli, ``Physics 
with nonperturbative quantum gravity: radiation from a quantum black 
hole'' {\em Gen. Rel. Grav.} {\bf 28} (1996) 1293-1299; 
Online Preprint Archive: {\em http://xxx.lanl.gov/abs/gr-qc/9603064}.

\bibitem{TQFT} M. F. Atiyah, {\em Publ. Math. IHES} {\bf 68} (1989) 
175. \\
E. Witten, {\em Com. Math. Phys.} {\bf 121} (1989) 351.

\end{thebibliography}
\end{document}